\documentclass[10pt]{article}

\usepackage{amsmath}

\usepackage{array}

\usepackage{appendix}

\usepackage{tocloft}                   

\usepackage{graphicx}

\usepackage{amsfonts}

\usepackage{amssymb}

\usepackage{mathrsfs}

\usepackage{yfonts}

\usepackage{euscript}

\usepackage{centernot}                 

\usepackage{ifsym}                     

\usepackage{upgreek}

\usepackage{mathtools}

\usepackage{color}

\usepackage{slantsc}
\usepackage{calligra}

\usepackage{bbold}          

\usepackage[T1]{fontenc}

\usepackage{epsf}

\usepackage{latexsym}

\usepackage{tipa}

\usepackage{makeidx}

\makeindex

\def\b{\begin{equation}}

\def\e{\begin{equation}}

\def\be{\begin{equation}}              

\def\ee{\end{equation}}

\def\beq{\begin{equation}}

\def\eeq{\end{equation}}

\def\bea{\begin{eqnarray}}

\def\eea{\end{eqnarray}}

\def\m{\mbox{ }}

\def\mma {\m , \m \m }

\def\!{\hspace{-1.6667em}}

\def\Proof{{\n{\u{Proof}}} \m}

\def\c{\cite}

\def\l{\label}

\def\r{\ref}

\def\n{\noindent}

\def\f{\footnote}

\def\u{\underline}

\def\w{\widetilde}

\def\s{\stackrel}

\def\biC{\mbox{\boldmath$C$}}              

\def\mA{\mbox{A}}

\def\mC{\mbox{C}}                        

\def\mD{\mbox{D}}                        

\def\mE{\mbox{E}}                        

\def\mI{\mbox{I}}                        

\def\mN{\mbox{N}} 

\def\mO{\mbox{O}}

\def\mP{\mbox{P}}

\def\mR{\mbox{R}}                        

\def\ma{\mbox{a}}

\def\mb{\mbox{b}}

\def\me{\mbox{e}}

\def\mf{\mbox{f}}

\def\ml{\mbox{l}}

\def\mo{\mbox{o}}

\def\mp{\mbox{p}}

\def\mr{\mbox{r}}

\def\uR{\u{R}}

\def\uq{\u{\mbox{q}}} 

\def\ur{\u{\mbox{r}}}

\def\urho{{\u{\rho}}}

\def\bA{\mbox{\bf A}}

\def\bE{\mbox{\bf E}}

\def\bS{\mbox{\bf S}}

\def\fI{\mbox{\sffamily I}}

\def\sg{\mbox{\scriptsize g}}

\def\sp{\mbox{\scriptsize p}}

\def\sq{\mbox{\scriptsize q}}

\def\sC{\mbox{\scriptsize C}}

\def\sE{\mbox{\scriptsize E}}

\def\sS{\mbox{\scriptsize S}}

\def\sfI{\mbox{\sffamily{\scriptsize I}}}      

\def\tcR{\mbox{\tiny ${\cal R}$}}

\def\es{\m = \m}

\def\:={\m := \m}

\def\=:{\m =: \m}

\def\cr{\mbox{\scriptsize{\bf $\m \times \m$}}}

\def\sumi2{\sum\mbox{}_{\mbox{}_{\mbox{\scriptsize $i$=1}}}^2}

\def\sumi3{\sum\mbox{}_{\mbox{}_{\mbox{\scriptsize $i$=1}}}^3}

\def\sumABcycles3{\sum\mbox{}_{\mbox{}_{\mbox{\scriptsize cycles $A,B$=1}}}^{3}}

\def\sumCDcycles3{\sum\mbox{}_{\mbox{}_{\mbox{\scriptsize cycles $C,D$=1}}}^{3}}

\def\sumj3{\sum\mbox{}_{\mbox{}_{\mbox{\scriptsize $j$=1}}}^3}

\def\sumk3{\sum\mbox{}_{\mbox{}_{\mbox{\scriptsize $k$=1}}}^3}

\def\prodiA1{\prod\mbox{}_{\mbox{}_{\mbox{\scriptsize $i$=1}}}^{A - 1}}

\def\d{\textrm{d}}                                                  

\def\pa{\partial}                                                   

\def\sin{\mbox{sin}}

\def\tan{\mbox{tan}}

\def\FrI{\mbox{$\mathfrak{I}$}}                                

\def\FrT{\mathfrak{T}}                                         
                                                       
\def\FrX{\mathfrak{X}}                                         
\def\FrS{\mbox{\Large $\mathfrak{s}$}}                         
\def\sFrS{\mbox{\large$\mathfrak{s}$}}                         
   
\def\tFrS{\mbox{\footnotesize$\mathfrak{s}$}} 
\def\FrU{\mbox{$\mathfrak{U}$}}                                
\def\sFrU{\mbox{\scriptsize$\mathfrak{U}$}}                    

\def\FrV{\mbox{$\mathfrak{V}$}}                                
\def\sFrV{\mbox{\scriptsize$\mathfrak{V}$}}                    

\def\FrW{\mbox{$\mathfrak{W}$}}                                
\def\FrM{\mbox{$\mathfrak{M}$}}                                
\def\lFrg{\mbox{\Large$\mathfrak{g}$}}                         
\def\FrT{\mbox{\boldmath$\mathfrak{T}$}}                       
\def\Hilb{\mbox{{\boldmath$\mathfrak{H}$}ilb}}                 
\def\FrQ{\mbox{\Large $\mathfrak{q}$}}                               
\def\bFrL{\mbox{\boldmath$\mathfrak{L}$}}                            

\def\Phase{\mbox{{\boldmath$\mathfrak{P}$}hase}}                     

\def\bFrR{\mbox{\boldmath$\mathfrak{R}$}}                            
\def\Rig-Phase{\bFrR\mbox{ig-}\Phase}                                
                                                                													   
\def\FrP{\mbox{\Large $\mathfrak{p}$}}                                 
\def\FrR{\mbox{\boldmath$\mathfrak{R}$}}                             

\def\1mat{\u{\u{1}}}                                                 

\def\Leib{\bFrL\mbox{eib}}                                           

\def\Positive-Modespace{\mbox{{\boldmath$\mathfrak{M}$}odespace$^+$}}

\def\POSITIVE-MODESPACE{\mbox{{\boldmath$\mathfrak{M}$}ODESPACE$^+$}}

\def\scR{\mbox{\scriptsize ${\cal R}$}}

\def\FrO{\mbox{$\mathfrak{O}$}}                                      

\def\Top{\FrT\mo\mp}

\def\lattice{\mbox{\bf\Large$\mathfrak{L}$}}                                      

\def\Kin-Hilb{\mbox{{\boldmath$\mathfrak{K}$}in-\Hilb}}                     

\def\Mid-Hilb{\mbox{{\boldmath$\mathfrak{M}$}id-\Hilb}}                     

\def\Dyn-Hilb{\mbox{{\boldmath$\mathfrak{D}$}yn-\Hilb}}                     

\def\5Star{\mbox{\Large$\star$}}              

\def\Frr{\mbox{$\mathfrak{r}$}} 

\def\Frs{\mbox{$\mathfrak{s}$}}

\begin{document}

\begin{titlepage}

\begin{center}

\Large{\bf Monopoles of Twelve Types in 3-Body Problems} \normalsize

\vspace{0.1in}

{\large \bf Edward Anderson$^*$}

\vspace{.2in}

\end{center}

\begin{abstract}

We consider twelve different ways of modelling the 3-body problem in dimension $\geq 2$. 
We show that a different type of monopole is realized in each's relational space: a type of reduced configuration space.
8 cases occur in 2-$d$, and 4 distinct ones in 3-$d$; these reflect counts of non-equivalent subgroup actions of $S_3 \times C_2$ and $S_3$ respectively. 
The $S_3$ acts on particle labels; the extra $C_2$ corresponds to the purely 2-$d$ option of whether or not to identify mirror images.  
The non-equivalent realization is due to a suite of subgroup, orbit space and stratification features.

\m

Our 2-$d$ monopoles include 4 known ones: a realization of Dirac's monopole in relational space rather than its more habitual setting of space, 
                                                            the 2-$d$ version of Iwai's monopole, 
													    and indistinguishable particle monopoles with and without mirror image identification.   
The 4 new ones are indistinguishable under a 2-particle label switch or under even permutations, in each case with optional mirror image identification.   
Our 4 3-$d$ monopoles are 2 known ones: the actual Iwai monopole and its already-announced indistinguishable-particles counterpart, 
                      and 2 new ones: the two-particle label switch and even permutation cases.  
All 4 3-$d$ cases are stratified.  
The three even-permutation cases are orbifolds, two with boundary, the 3-$d$ case's boundary constituting a separate stratum, giving a stratified orbifold.  
We document each of the 12 cases' underlying shape space and relational space, 
and each monopole's Hopf mathematics, 
global-section versus topological quantization dichotomy, 
Dirac string positioning, 
and Chern integral concordance with topological contributions form of Gauss--Bonnet Theorem. 
														
\end{abstract}

\n PACS: 04.20.Cv, 02.40.-k, Physics keywords: Monopoles, Background Independence, configuration spaces, 3-body problem, (in)distinguishability. 

\m

\n Mathematics keywords: Fibre and General Bundles, Hopf map, Applied Geometry, Applied Topology, Shape Theory, Shape Statistics, orbifolds. 

\vspace{0.1in}
  
\n $^*$ Dr.E.Anderson.Maths.Physics@protonmail.com

\section{Introduction}\l{Intro-III}

\n The more usual arena in which the well-known {\it Dirac monopole} \c{DirMon, Dirac48, Nakahara, Frankel} is theoretically realized is in physical space.
It is moreover also realized in a certain 3-body problem's configuration space, which furthermore admits a number of variants. 

\m

\n The current paper considers 3-body configurations; these are triangles in the sense of constellations of 3 points.   
It is these points, rather than the associable edges or lamina that are accorded a priori significance.  
Depending on the level of treatment (explained in Sec 2), triangles in this sense include all, or almost all, degenerate configurations: collinearities and collisions.  
Our approach moreover considers quotienting out translations, rotations, and, for some purposes, dilations.  
Quotienting out all three gives a theory of similarity shape, whereas retaining the dilations gives a theory of Euclidean shape-and-scale.  
Both formulations pass through use of relative separations (which for triangles are the same as sides) 
to equable use of sides and medians (under the assumption of equal-mass particles), alongside the angles between sides and medians. 
Both formulations make further use of Jacobi mass-weighted variables \cite{Marchal} 
and side-to-median ratios \cite{Ineq}, with the scaled case's (Jacobi mass-weighted) natural scale variables 
being the total moment of inertia and its square root.  

\m

\n The shape space $\FrS(3, 2)$ of pure similarity shape triangles turns out to be a sphere $\mathbb{S}^2$ \cite{Kendall84, Kendall89}, 
whereas the relational space $\FrR(3, 2)$ of Euclidean scaled triangles is the cone thereover \cite{Cones, FileR}, 
These and further such results -- Shape Theory -- are outlined in Sec 2, concentrating on spatially 1- and 2-$d$ results.  
This involves passing from the unreduced configuration space of particle position vectors to, 
firstly, relative space in the obvious Lagrange--Jacobi relative position vectors sense \c{Marchal}, by quotienting out translations.  
Secondly, to the preshape space in David Kendall's sense of unit moment of inertia relative position vectors \c{Kendall84, Kendall89, Kendall}, by quotienting out dilations. 
Thirdly, to Kendall's shape space by additionally quotienting out rotations; in 2-$d$ this is elegantly described by Hopf mathematics \c{Hopf, Nakahara, Frankel}.
Kendall's work remains rather more familiar in the Shape Statistics literature \c{Small, Kendall, Bhatta, DM16, PE16}; 
see also e.g.\ \c{LR95, LR97, Montgomery, HS, AF, +Tri, FileR, QuadI, APoT2, QuadII, AKendall, ABeables, ABeables2, MIT, ABeables3, ABook, 
I, II, III, A-Pillow, 2-Herons, A-Pillow-2, Ineq} for related work in other fields, including Mechanics and modelling some aspects of General Relativity's Background Independence.
Fourthly, by taking the cone over shape space, we pass to relational space, 
meant in the sense of the absolute versus relational debate \c{ABook} which goes back (at least) to Newton versus Leibniz. 

\end{titlepage}

\n On the one hand, 3 particles in 1-$d$ does not support any monopoles, and so plays no further role in this paper (see \cite{I} for a recent review).   
On the other hand, 3 particles in 3-$d$ (or higher) has some subtle differences with 3 particles in 2-$d$. 
Due to this, Secs 3 to 4 consider the 2-$d$ case first, 
and then Secs 5 and 6 consider these differences alongside the distinct theory of 3-$d$ 3-particles (scale-and-)shape space and their monopoles.
Sec 3 considers modelling variety in 2-$d$ 3-body problems from quotienting out discrete transformations: 
(partial) particle label indistinguishability and/or mirror image identification. 
This variety rests upon the groups and lattices considerations of Appendices A and B, and in turn sources a dual diversity of monopoles as presented in Sec 4.   

\m 

\n Kendall \c{Kendall84, Kendall89} already considered the maximal case of this discrete quotienting, 
giving a shape space `fundamental cell' that he termed the spherical blackboard.
This has special status as the bottom element of the bounded lattice of discrete quotientings. 
It is also the basic tile with which shape space can be tessellated \cite{Kendall89}, an approach \cite{+Tri, FileR, III} which is useful in the current paper.  
The opposing top element is just the shape sphere itself.
We term the bottom element in the general $N$-particle $d$-dimensional case the {\it Leibniz space} due to its implementation of Leibniz's `Identity of Indiscernibles' \cite{L}.  

\m

\n The variety of similarity shape spaces of triangles from quotienting out discrete transformations turns out to give a lattice of size 8 in 2-$d$.
This corresponds to whether mirror image identification is carried out, and whether particle labels are identified, including identifying 2 of the 3 labels 
and identification under even permutations alone.    
Each of the above eightfold produces its own type of monopole, giving the bounded lattice of monopoles presented in Sec 5.  
The top    monopole in the lattice is the relational space realization of the familiar Dirac monopole, whereas 
the bottom monopole is the `Leibniz--Kendall' monopole described in \cite{III}.  
2 of our other 6 2-$d$ monopoles were already mentioned in \cite{III} whereas the other 4 are new to the current paper.

\m 

\n In 3-$d$, mirror image identification becomes obligatory, by freedom to rotate through the third dimension. 
This cuts down the number of discrete quotients to 4. 
Prima facie, these look very similar to our mirror image identified 4 2-$d$ configuration spaces, 
whose top element is the shape hemisphere and whose bottom element is the `same' Leibniz space as for 2-$d$.  
However, 3-$d$ involves quotienting out $SO(3)$, which unlike its 2-$d$ counterpart $SO(2)$, has a nontrivial continuous subgroup: 
\be 
SO(2 \m) < \m SO(3) \m . 
\ee 
A consequence of this is that in 3-$d$, some configurations -- the collinear ones -- are only acted upon by the $SO(2)$ subgroup. 
Thereby, these form a separate stratum. 
So the 3-$d$ top shape space is, more specifically, a hemisphere with a mathematically distinct edge -- the equator of colinearity -- 
whereas the 2-$d$ shape hemisphere's edge is not distinct in this manner. 
This distinction moreover pervades all 4 cases, by which we are dealing overall with a suite of 8 + 4 = 12 distinct shape spaces, 
and thus of 12 relational spaces and 12 species of monopoles.  
We show how each case's relational space cone over shape space decomposes into strata.  
The 3-$d$ lattice's top monopole in 3-$d$ is moreover the Iwai monopole \cite{Iwai87}, 
with corresponding bottom monopole the `Leibniz-Kendall stratified monopole' described in \cite{III}. 
The other two 3-$d$ 3-body problem monopoles are new to the current paper.  

\m 

\n Study of these monopoles benefits from identifying a key underlying mathematical structure: {\it Hopf-type bundles}. 
For the Dirac monopole itself, it is the usual Hopf fibre bundle \c{Hopf, Nakahara, Frankel} that is realized, corresponding to full-range Hopf coordinates.  
In the variants, one usually needs a general bundle \c{Isham, Husemoller} rather than a fibre bundle version, 
though the {\it Hopf coordinates} themselves remain ubiquitous (if with different ranges).
For Iwai's monopole (or its 2-$d$ counterpart with non-distinct edge), a half-space range of the Hopf coordinates are realized (without or with the edge value included):  
{\it Dragt coordinates} \c{Dragt}. 
In other cases a distinct 1/3 or 1/6 range occurs, the former with topological identification to form an orbifold stucture.   
Appendix E extends Sec 2's outline of the Hopf map to bundle-theoretic considerations.  

\m

\n Secs 4 and 6 furthermore indicate where each monopole's Dirac strings can be placed, alongside which admit global sections 
and which necessitate transition functions between local sections. 
The latter incur topological quantization conditions (as outlined in Appendix C) whereas the former do not. 
We present suitable ways of placing each monopole's Dirac string. 
We also compute each monopole's Chern integral \c{Frankel}, checking that this matches up with the total topological contribution 
-- which has Euler characteristic and external angle sum edge-contributions -- by which a form of Gauss--Bonnet Theorem (explained in Appendix D) holds.  

\m 

\n 3 of the current paper's new 3-body problem monopoles are orbifolds: one without boundary with two conical singularities, and two with boundary and one conical singularity, 
differing as to whether said boundary is a separate stratum.  
Thus both stratification and orbifolds are present in the current paper's monopoles, including one instance in which both features coexist.  
Our Conclusion (Sec 7) includes mention of $N$-a-gon extensions, 
                                           quantization applications, 
									   and of the varying levels of complexity of stratification found so far in Shape Theory (supported by Appendix F); 
the current paper's being particularly mild, whereas \cite{GT09, KKH16}'s is particularly strong.

\vspace{10in}

\section{Continuous quotients of $N$- and $3$-particle configuration spaces}

\subsection{Topological triangles and their spaces}\label{TTS}
%
{            \begin{figure}[!ht]
\centering
\includegraphics[width=0.40\textwidth]{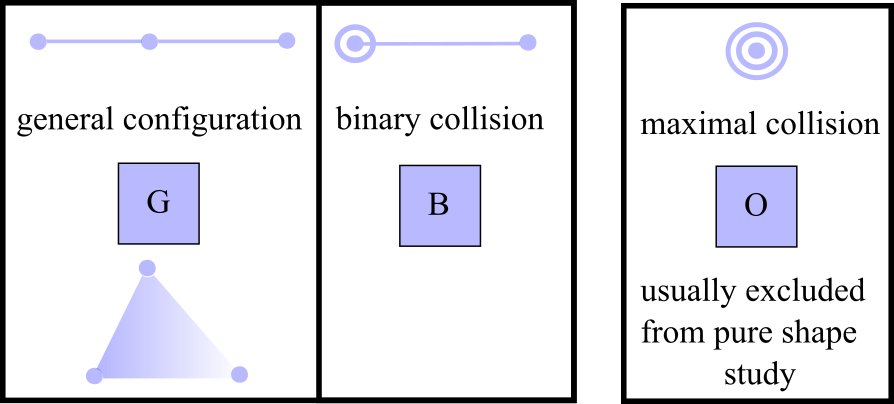}
\caption[Text der im Bilderverzeichnis auftaucht]{        \footnotesize{Topological classes of configurations for 3 particles in 1-$d$:  
the collisionless generic configuration G which covers both triangular and collinear metrically distinguished subcases, 
the binary collisions B, and 
the maximal collision O, which is usually excluded from pure shape study and even from some shape-and-scale considerations. 
We depict topological shapes in   pale blue to render them immediately distinguishable from 
    topological shapes' spaces (bright blue), 
    metric shapes              (grey) 
and metric shape spaces        (black).} }
\l{(3, 2)-Top-Shapes} \end{figure}          }
%
\n{\bf Proposition 1} \cite{III} There are three topological types of 3-point shape as per Fig \r{(3, 2)-Top-Shapes}.  

\m

\n{\bf Remark 1}
\be
\mbox{\#(G)} = 1 \m : 
\ee
the generic `rubber triangle' (including collinear cases but excluding binary or maximal collisions) can be deformed from all labellings to all other labellings. 
Without discrete identifications,
\be 
\mbox{\#(B)} = \mbox{(ways of leaving one particle out)} 
             = 3                                            \m .  
\ee
The above represent two salient differences with the 1-$d$ case.  

\m

\n{\bf Proposition 1} \cite{III} For distinguishably labelled points, the topological shape space is
\be
\Top\mbox{-}\FrS(3, 2) = \mbox{Claw} \m :
\ee
the 4-vertex {\it claw graph} with cluster-labelled `talons' as per Fig \r{S(3, 2)-Top-Dist}.     
%
{            \begin{figure}[!ht]
\centering
\includegraphics[width=0.36\textwidth]{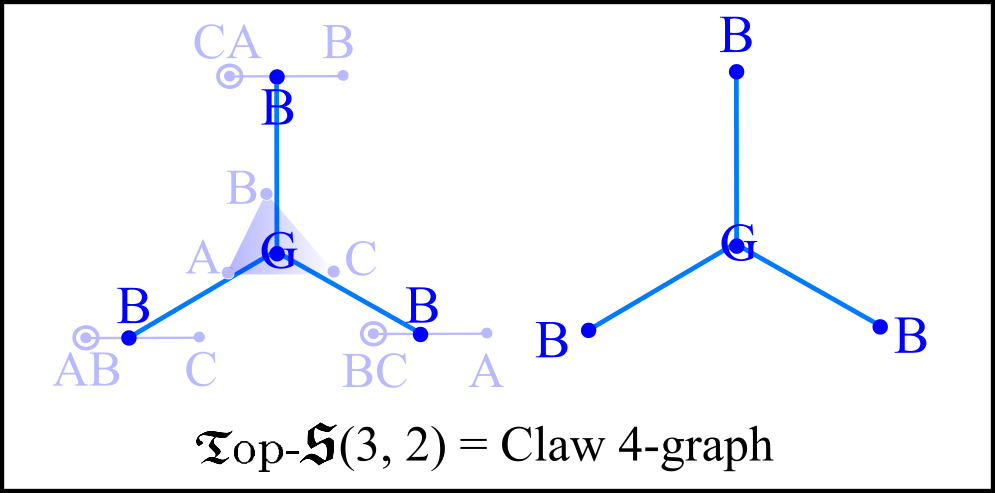}
\caption[Text der im Bilderverzeichnis auftaucht]{        \footnotesize{Topological shape space for topological triangles with distinctly-labelled vertices. 
The left-hand-side version illustrates which topological triangles correspond to which vertices.
We use bright blue so as to immediately pick out topological shape spaces. } }
\l{S(3, 2)-Top-Dist} \end{figure}          }
 
\n More generally, spaces of topological shapes are always graphs; as \cite{II} starts to show, these are more complicated in 1-$d$ than in higher-$d$ (for $N \geq 4$).

\subsection{Metric-level triangles and other constellations}
%
{            \begin{figure}[!ht]
\centering
\includegraphics[width=1.0\textwidth]{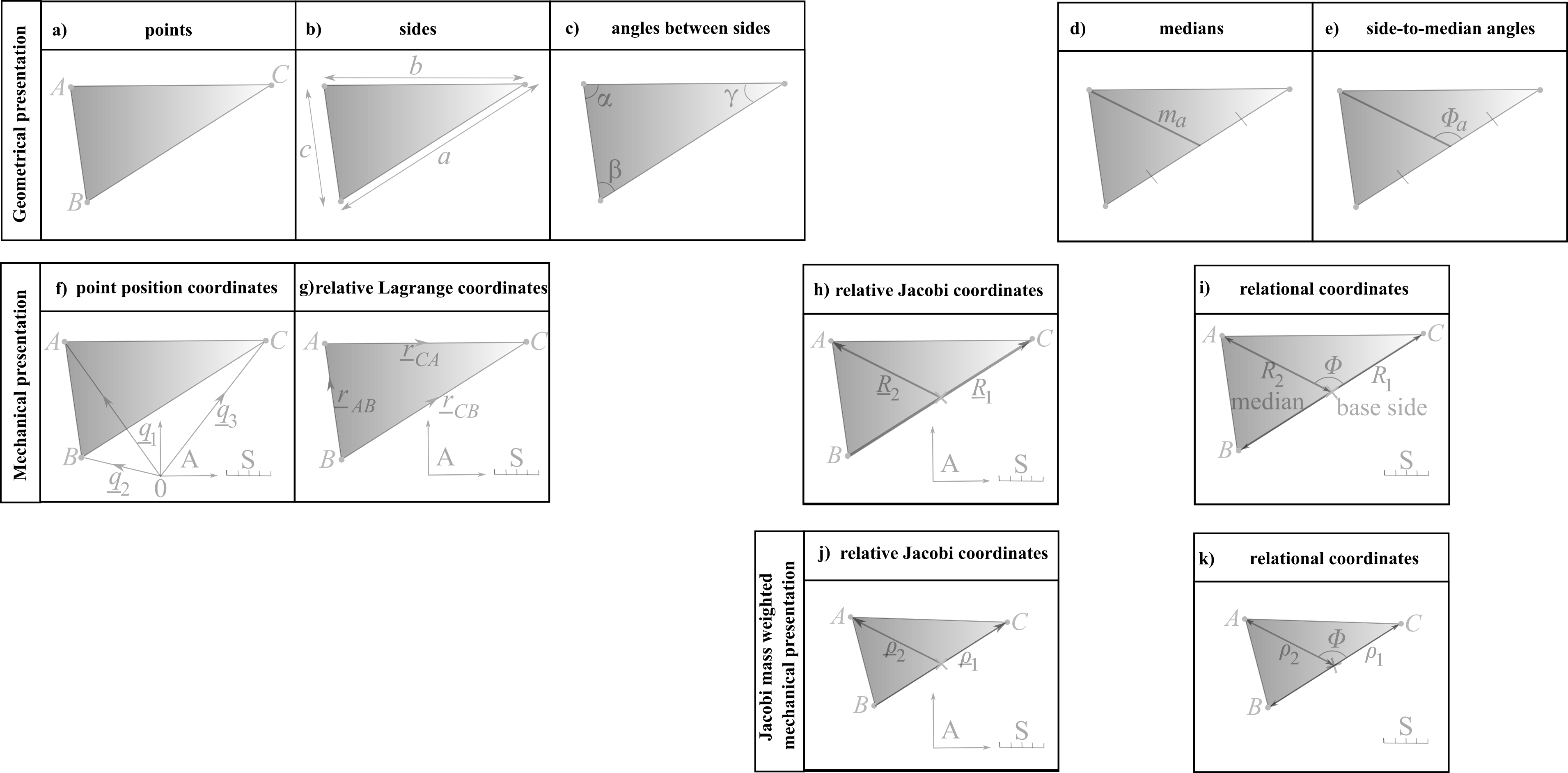}
\caption[Text der im Bilderverzeichnis auftaucht]{        \footnotesize{Geometrical and mechanical notation for the general triangle.
I use grey for metric triangles in space.} }
\l{Tri-q-r-rho-relatio}\end{figure}            }
%
\n{\bf Remark 1} Fig \r{Tri-q-r-rho-relatio}.a)-c)'s standard geometrical description of the general triangle can be rephrased  
in terms of Fig \r{Tri-q-r-rho-relatio}.f)'s {\it position coordinates} $\uq_I$ for the triangle's vertices -- now viewed as particles -- 
relative to an absolute origin 0, axes A and scale S.

\m

\n{\bf Structure 1} Let us next pass to {\it relative Lagrangian coordinates} (Fig \r{Tri-q-r-rho-relatio}.e), consisting of particle separation vectors 
\be
\ur_{IJ} = \uq_J - \uq_I  \m .  
\ee 
\n{\bf Remark 2} Using these amounts to discarding the absolute origin 0.  
Their magnitudes return the side-lengths and the angles between these vectors return Fig \r{Tri-q-r-rho-relatio}.c)'s angles between sides.  

\m

\n{\bf Remark 3} Moreover, not all of the relative Lagrange coordinates are independent. 
This can be circumvented by picking a basis of two of them.  
This however leaves the inertia quadric in non-diagonal form.  

\m

\n{\bf Structure 2} Diagonalizing the inertia quadric leads us on to {\it relative Jacobi coordinates}.
These are a more general notion of particle {\sl cluster} separation vectors, i.e.\  particle subsystem centre-of-mass separations.   
They include the relative Lagrange coordinates as a special subcase: the one for which the particle subsystems at both ends consist of one particle each, 
by viewing this particle as the location of its own centre of mass.
In particular, in the case of a triangular configuration, 
\be
\uR_1  =   \uq_C - \uq_B                    \mma 
\uR_2 \es  \uq_A - \frac{\uq_B + \uq_C}{2}  \m , 
\l{R-def}
\ee 
or cycles thereof for the other two choices of 2-particle cluster.  
These have associated cluster masses (conceptually reduced masses), respectively  
\be 
\mu_1 \es  \frac{1}{2}    \mma  
\mu_2 \es  \frac{2}{3}    \m . 
\ee 
\n{\bf Structure 3} One can furthermore pass to {\it mass-weighted relative Jacobi coordinates} (Fig \r{Tri-q-r-rho-relatio}.j)
\be 
\urho_i := \sqrt{\mu_i}\uR_i  \m \m  (i = 1 \mma  2)  \m .   
\l{rho-R}
\ee 
These leave the inertia quadric's matrix as the identity, and furthermore turn out to be shape-theoretically convenient to work with.

\m

\n{\bf Structure 4} Passing to mass-weighted relative Jacobi coordinates $\urho_a$ does not however have any effect as regards removing the absolute axes A or scale S, 
since these remain vectorial, and thus made reference to absolute directions and absolute magnitudes. 

\m

\n Reference to absolute axes A is moreover removed by restricting attention (Fig \r{Tri-q-r-rho-relatio}.e) to the following triple.

\m

\n i)-ii) The {\it relative Jacobi magnitudes} $\rho_a$ ($a = 1$, 2).   
Comparing Fig \r{Tri-q-r-rho-relatio}.a) and d) with the mass-unweighted version of Fig \r{Tri-q-r-rho-relatio}.h), these are a base side and the corresponding median respectively.
 
\m 
 
\n iii) The {\it relative Jacobi angle} between the $\urho_a$, 
\be 
\Phi  \es  \mbox{arccos}    \left(    \frac{\urho_1\cdot\urho_2}{\rho_1\rho_2}  \right)
      \es  \mbox{arccos}    \left(    \frac{\sqrt{\mu_1} \uR_1 \cdot \sqrt{\mu_2} \uR_2}{ \sqrt{\mu_1} R_1 \sqrt{\mu_2} R_2} \right) 
      \es  \mbox{arccos}    \left(    \frac{\uR_1 \cdot \uR_2}{ R_1 R_2}     \right)                                                   \m . 
\l{Swiss}
\ee  
This is a `{\it Swiss-army-knife relative angle}' as per Fig 3.i) or k).    
Note that the above demonstrates this angle to be unaltered by mass-weighting, 
by which it is identical to the geometrical side-to-median angle $\Phi_a$ of Fig \r{Tri-q-r-rho-relatio}.e)

\m

\n Together, ($\rho_1, \rho_2, \Phi$) constitute {\it relational scale-and-shape data} for the triangle.

\m

\n {\bf Structure 4} Finally, reference to absolute scale S is removed by continuing to consider the relative Jacobi angle, but now just alongside 
the {\it ratio of Jacobi magnitudes},  
\be 
{\cal R}  \:=  \frac{\rho_2}{\rho_1}  \m .
\l{cal-R-def}
\ee 
In \c{III} we also ascertained that this provides a further similarity condition for triangles and, accordingly, serves as data from which 
the `primary' triangle's features of Fig 1.a) modulo scale -- side-to-side relative angles and side-length ratios -- can be abstracted.

\subsection{Metric-level configuration spaces for triangles and other constellations}\l{Met-Config}
%
{            \begin{figure}[!ht]
\centering
\includegraphics[width=1.0\textwidth]{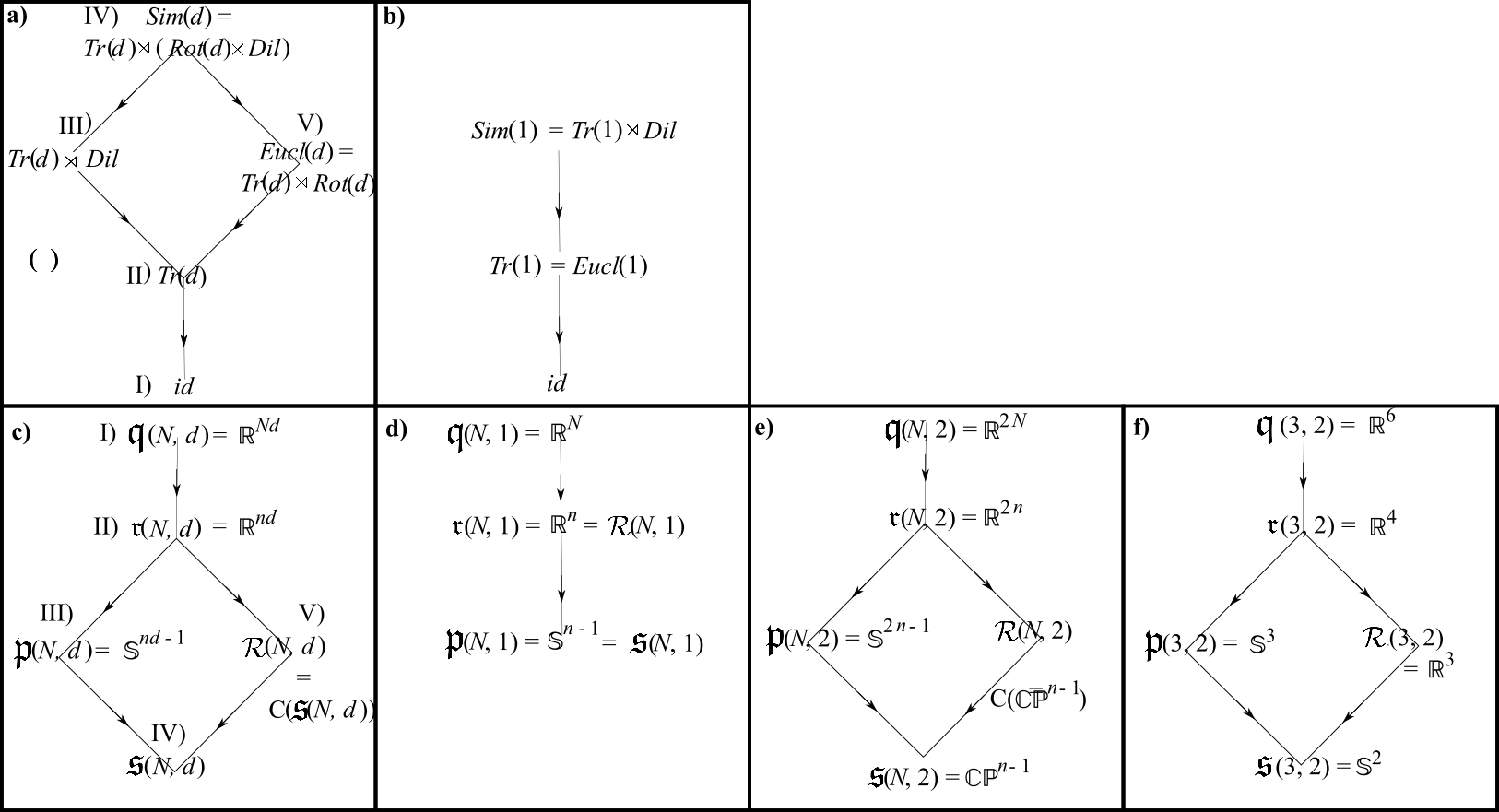}
\caption[Text der im Bilderverzeichnis auftaucht]{        \footnotesize{a) Lattice of 5 subgroups of the similarity group $Sim(d)$, with simplified version in 1-$d$ in b). 
c) is the corresponding dual lattice of configuration space quotients, 
specializing to dimension 1 in d), 
2 in e), 
and furthermore to the 3 particles in 2-$d$ of triangleland in f).} }
\l{Lattice-of-5}\end{figure}            }

\n We consider here certain particularly meaningful subgroups of the similarity group $Sim(d)$, which form the lattice displayed in Fig \ref{Lattice-of-5}.  

\m 

\n{\bf Structure I)} At the level of configuration spaces, the space of the $\uq_I$ for arbitrary particle number $N$ and dimension $d$ is straightforwardly 
\be 
\FrQ(N, d) =  \{\mathbb{R}^d\}^N 
           =    \mathbb{R}^{N \, d}  \m . 
\l{Q(N, d)}
\ee 
Let us next follow through what happens to this as one removes the absolute origin 0, axes A and scale S \c{Kendall84, Kendall, FileR}. 

\m

\n{\bf Structure II)} Quotienting out by $Tr(d)$ -- the $d$-dimensional translations -- 
we arrive at the configuration space of independent $\ur_{IJ}$, $\uR_i$ or $\urho_i$: {\it relative space} 
\be 
\Frr(N, d) \mbox{ } := \mbox{ } \frac{\FrQ(N, d)}{Tr(d)} 
           \mbox{ }  = \mbox{ } \frac{\mathbb{R}^{N \, d}}{\mathbb{R}^d} 
		   \mbox{ }  = \mbox{ } \mathbb{R}^{n \, d}                       \m , 
\l{r(N, d)}
\ee
where we have also defined the {\it independent relative particle separation number} $n := N - 1$.

\m

\n{\bf Structure III)} It is next useful to consider furthermore quotienting out the dilation group $Dil$ to render absolute scale S irrelevant. 
This yields Kendall's {\it preshape space}  
\be 
\FrP(N, d) \:=  \frac{\Frr(N, d)}{Dil} 
           \es  \frac{\mathbb{R}^{n \, d}}{\mathbb{R}_+} 
           \es  \mathbb{S}^{n \, d - 1}                   \m :   		   
\l{P(N, d)}		   
\ee
the ($n \, d - 1$)-dimensional sphere, which is the configuration space of ratios of independent components of $\ur_{IJ}$, $\uR_i$ or $\urho_i$.
This moreover carries the standard hyperspherical metric.

\m

\n{\bf Structure IV)} Quotienting out as well by $Rot(d)$ -- the $d$-dimensional rotations -- 
and so overall by $Sim(d)$: the $d$-dimensional similarity group of translations, rotations and dilations. 
we arrive at Kendall's {\it shape space} \cite{Kendall84, Kendall89, Kendall} is 
\be 
\FrS(N, d) \:=  \frac{\FrQ(N, d)}{Sim(d)} 
           \es  \frac{\Frr(N, d)}{Rot(d) \times Dil} 
		   \es  \frac{\FrP(N, d)}{Rot(d)}                 \m . 
\l{S(N, d)}		   
\ee 
\n{\bf Structure 1} Subsequent analysis picks up $d$ and $N$ dependence.
For $d = 1$, there are not continuous rotations to remove, so 
\be 
\FrS(N, 1) = \FrP(N, 1) 
           = \mathbb{S}^{n - 1}  \m . 
\ee 
For $d = 2$, 
\be
Rot(2) = SO(2) 
       = U(1) 
	   = \mathbb{S}^1  \m , 
\ee 
and \c{Kendall84} 
\be 
\FrS(N, 2) \es  \frac{ \mathbb{S}^{n \, d - 1}}{\mathbb{S}^1} 
           \es  \mathbb{CP}^{n - 1}                            \m , 
\l{S(N, 2)}		   
\ee 
which is most readily arrived at by the generalization of the Hopf map \c{Hopf} to odd-dimensional spheres, 
\be
{\cal H}_{\sS} \m : \m \m \mathbb{S}^{2 \, k + 1} \longrightarrow \mathbb{CP}^k  \m . 
\ee 
2-$d$ shape space moreover carries the natural Fubini--Study metric \c{Kendall84, Kendall}. 
This result more than suffices for the current paper; indeed, for $N = 3$ -- {\it triangleland} -- the Hopf map itself, 
\be
{\cal H}_{\sS} \m : \m \m  \mathbb{S}^3 \longrightarrow \mathbb{S}^2 \m ( \m = \mathbb{CP}^1 \m )            \m ,
\ee 
suffices. 
The Fubini--Study metric moreover simplifies in this $N = 3$ case, giving in plane-polar coordinates  
\be 
\d s^2  \es  4  \, \frac{\d {\cal R}^2 + {\cal R}^2\d\Phi^2}{  (  1 + {\cal R}^2  )^2  }  \m .   
\ee 
This can in turn be recognized as the standard spherical metric in stereographic coordinates. 
Thus, in the Shape Theory of triangles, the ratio of Jacobi magnitudes ${\cal R}$ of eq. (\r{cal-R-def}) plays the geometric role of stereographic radius of the shape sphere. 

\m
 
\n The Swiss-army-knife relative angle $\Phi$ of the shape-in-space thus plays the role of polar angle on the shape sphere. 

\m

\n The venerable substitution 
\be 
{\cal R} = \tan \, \frac{\Theta}{2}  \m  
\l{RTheta}
\ee 
finally serves to convert the stereographic radius to the standard azimuthal coordinate $\Theta$ (about the U-axis), 
casting the shape sphere metric into the standard spherical metric form 
\be 
\d s^2 = \d {\Theta}^2 + \mbox{sin}^2 \Theta \, \d\Phi^2  \m . 
\label{Sphe-Met}
\ee
\n{\bf Structure 2} See Fig \r{S(3, 2)-Intro} for where some of the most qualitatively distinctive triangle shapes-in-space are realized in the triangleland shape space.
%
{            \begin{figure}[!ht]
\centering
\includegraphics[width=0.45\textwidth]{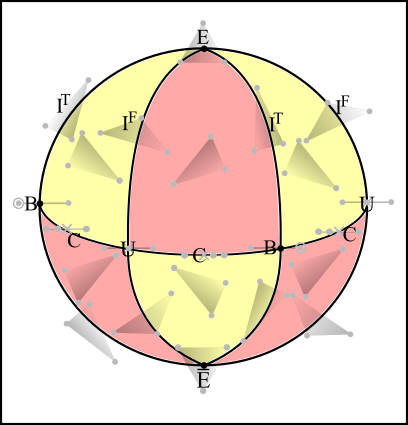}
\caption[Text der im Bilderverzeichnis auftaucht]{        \footnotesize{a) The triangleland shape sphere \c{Kendall89, +Tri, FileR, ABook, III}.  
Equilateral triangles E are at its poles, whereas collinear configurations C form its equator.
C separates hemispheres of clockwise and anticlockwise oriented triangles.  
There are 3 bimeridians of isoscelesness I corresponding to 3 labelling choices for the vertices.
T and F superscripts stand for `tall' and `flat'. 
$\mC \, \cap \, \mI$ gives 3 binary collisions B \c{Montgomery}, which are moreover topologically significant, and 3 uniform collinear shapes U.
Each of these triples lies on the equator, with each member at $\frac{2 \, \pi}{3}$, and the two triples at $\frac{\pi}{3}$ to each other.} }
\l{S(3, 2)-Intro} \end{figure}          }

\m

\n Note that the spherical coordinates we have found have U and B as their North and South poles.
There are moreover 3 clustering choices of such axes, realized at  $\frac{2 \, \pi}{3}$ angles to each other in the plane of collinearity.  
For some purposes, one would prefer to use the more distinguished and cluster-independent $\mE\overline{\mE}$ axis to define new spherical polar coordinates 
$\widetilde{\Theta}$, $\widetilde{\Phi}$ \c{FileR, III}.  
Working in terms of these is however more involved in terms of moving back and forth between shapes in space and points, curves and regions in the shape sphere.  
Moreover, we only need one set of spherical polars to establish the shapes-in-space to points-in-shape-space correspondence.  

\vspace{10in}

\n{\bf Structure V)} Finally, quotienting out by the Euclidean group $Eucl(d)$ of translations and rotations, we arrive at the {\it relational space} 
\be 
{\cal R}(N, d)  \:=  \frac{\FrQ(N, d)}{Eucl(d)} 
                \es  \frac{\Frr(N, d)}{Rot(d)}
\ee 

\m

\n{\bf Structure 3} A general result for the topological and geometrical form of this is that 
\be 
{\cal R}(N, d) = \mC(\FrS(N, d))  \m , 
\l{R(N, d)}
\ee 
for C the topological and metric coning construct.
In particular, for the 2-$d$, $N = 3$ triangleland, 
\be 
{\cal R}(3, 2) = \mC(\mathbb{S}^2) 
               = \mathbb{R}^3
\ee 
topologically (albeit it is not metrically flat, though it is conformally flat if the maximal collision O is omitted, see e.g. \c{FileR}).

\m

\n{\bf Structure 4} is the 2-$d$ shape-theoretic implementation of the extended generalized Hopf map, as laid out in Fig \r{Generalized-Hopf-Map}.   
%
{            \begin{figure}[!ht]
\centering
\includegraphics[width=0.75\textwidth]{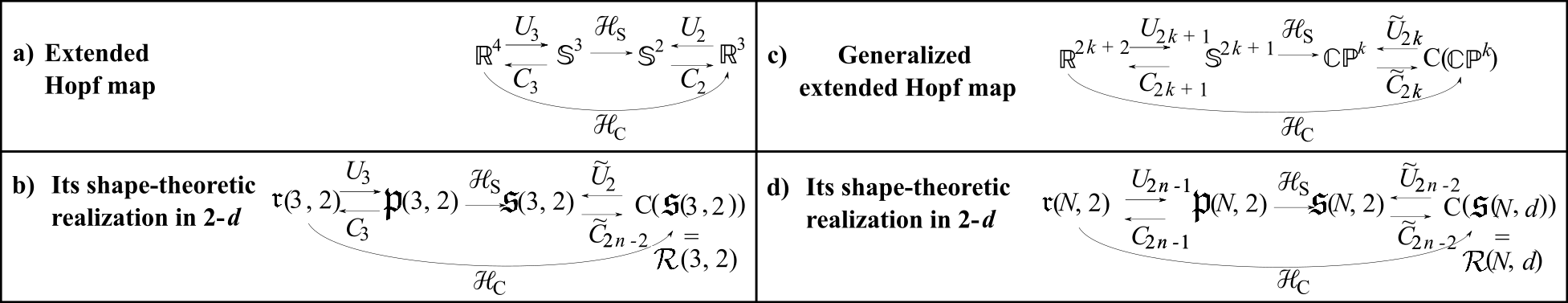}
\caption[Text der im Bilderverzeichnis auftaucht]{        \footnotesize{a) In the special case of the triangle, 
the extension involves unit maps $U$ and coning maps $C$.

\m 

\n b) Its shape-theoretic implementation using eqs (12), (13), (19) and (25). 

\m

\n c) In the general case, the extension involves unit maps $U$ and coning maps $C$, 
the tilded such now being other than between spheres and their ambient real spaces of dimension one higher. 

\m 

\n d) Its shape-theoretic implementation, for $k = n - 1$ and using eqs (\r{r(N, d)}), (\r{P(N, d)}), (\r{S(N, 2)}) and (\r{R(N, d)}). }  } 
\l{Generalized-Hopf-Map} \end{figure}           }

\n The extended Hopf map is 
${\cal H}_{\sC}$'s associated Cartesian coordinates bear (moving left) the following relation to the mass-weighted relative Jacobi interpretations 
and 3-body problem specific interpretations (moving right; see \cite{III, 2-Herons} for detailed interpretation). 
\be
2 \, \urho_1 \cdot \urho_2           \es  Hopf_x  
                                     \es  (\mbox{ anisoscelesness })  \m , 
\label{Hopf-x}
\ee 
\be
2 \, \{ \urho_1 \cr \urho_2 \}_3     \es  Hopf_y  
                                     \es  (\mbox{ 4 $\times$ area })  \m , 
\label{Hopf-y}
\ee 
\be
\urho_1\mbox{}^2 - \urho_2\mbox{}^2  \es  Hopf_z  
                                     \es  (\mbox{ ellipticity })       \m . 
\label{Hopf-z}
\ee 
These each have the coordinate range $(-\infty, \, \infty)$ and correspond to the untilded    $\Theta$ and     $\Phi$; 
                                                                                          $\w{\Theta}$ and $\w{\Phi}$ correspond to reversing $y$ and $z$'s status.

\section{Discrete quotients of shape spaces}\label{Disc-Quot}

\subsection{Action of $S_3 \times C_2$ on the shape space of triangles}

\n{\bf Remark 1} We consider quotienting out $S_3$, so as to model rendering the vertex labels indistinguishable. 
Note that this renders equivalent the different labellings of isosceles triangles I, as well as the corresponding left- and right-leaning triangles.  

\m

\n{\bf Remark 2} We separately quotient out by mirror image identification $C_2$. 
Note that this leaves the collinear configurations C -- which form the equator -- invariant while interchanging the hemispheres of clockwise and anticlockwise oriented triangles.  

\m 

\n{\bf Remark 3} Overall, this amounts to quotienting out by $S_3 \times C_2$, of order $6 \times 2 = 12$.

\m

\n{\bf Remark 4} In 3-$d$, mirror image identification is obligatory. 
Thus we are left with $S_3$ to quotient out.  
This is abstractly standard, so we can immediately read off its lattice of subgroups (Appendix A). 
In contrast, it is not immediately clear which of the abstractly-standard groups of order 12 $S_3 \times C_2$ is.  
We address this in Appendix B, so as to identify its lattice of subgroups as well.

\subsection{Consequent 3-chain of topological shape spaces}\label{3-Chain}
%
{            \begin{figure}[!ht]
\centering
\includegraphics[width=0.45\textwidth]{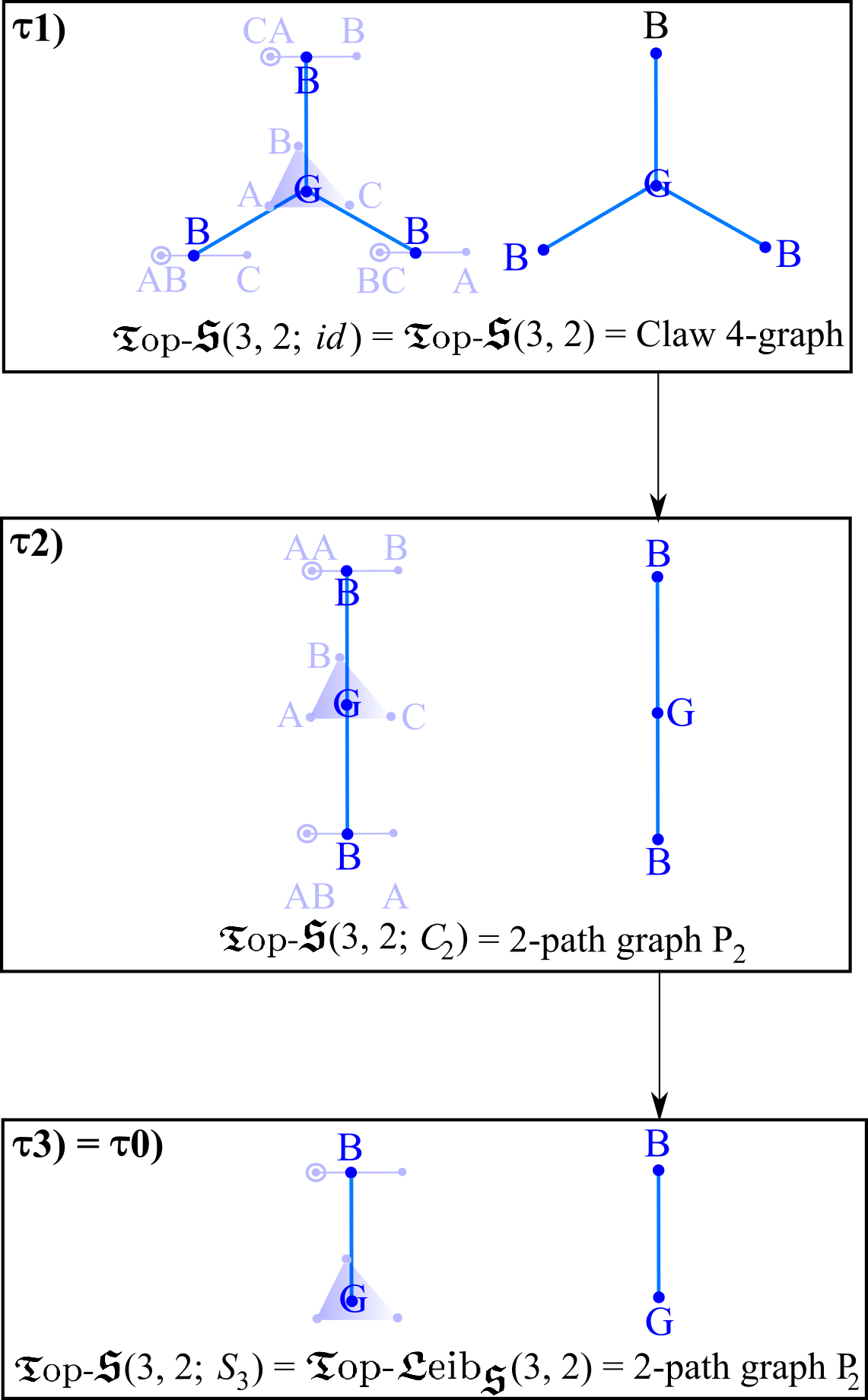}
\caption[Text der im Bilderverzeichnis auftaucht]{        \footnotesize{3-chain of topological shape spaces for topological triangles 
with the various distinctly-acting discrete quotientings implemented.
On the bottom rung, all equal labels is equivalent to {\sl no} labels. 
The $\tau$-labels stand for `topological' and the initially given numbers are the number of label equalities. 
The top rung's 1 label moreover doubling as lattice-theoretic top-1 label, the bottom rung's label is also reset to a bottom-0 as indicated.} }
\l{S(3, 2)-Top-Disc} \end{figure}          }

For topological shapes, there are 3 cases. 

\m 

\n{\bf Case $\tau$1)} 
\be 
\Top\mbox{-}\FrS(3, 2) := \Top\mbox{-}\FrS(3, 2; id) 
                        = \mbox{Claw}                 \m : \m \mbox{ 3-path graph (Fig \r{S(3, 2)-Top-Disc}.1)) }  \m .            
\ee 
\n{\bf Case $\tau$2)} For 2 indistinguishable particles (using $C_2$ for the remaining $C_2\mbox{-label}$ action)    
\be
\Top\mbox{-}\FrS(3, 2; C_2)  \es  \frac{\Top\mbox{-}\FrS(3, 2)}{C_2} 
                             \es  \frac{\mbox{Claw}}{S_2} 
							 \es  \mP_3                                \m : \m \mbox{ 3-path graph (Fig \r{S(3, 2)-Top-Disc}.2)) }  \m . 
\ee 
\n{\bf Case $\tau$3) = $\tau$0)} \cite{III} For indistinguishable particles,   
\be
\Top\mbox{-}\Leib_{\sFrS}(3, 2)  :=   \Top\mbox{-}\FrS(3, 2; S_3)  
                                \es  \frac{\Top\mbox{-}\FrS(3, 2)}{S_3}  
								\es  \frac{\mbox{Claw}}{S_2} 
							    \es  \mP_2                                \m : \m \mbox{ 2-path graph (Fig \r{S(3, 2)-Top-Disc}.0)) }  \m . 
\ee                          
\n{\bf Remark 1} The prefix $\Top$ is used to refer to spaces of {\sl topological} shapes.  
`$\Leib$' refers to {\it Leibniz space}: the maximal discrete quotient, by which it has by definition the status of bottom element in the lattice of discrete quotients.  
This is in opposition to the top element: the shape space with no discrete quotienting at all, $\FrS(N, d)$, which for triangles in 2-$d$ is the shape sphere.  
The second notation in the last case indicated that it is the bottom element of the chain.  
 
\m

\n{\bf Remark 2} All spaces of topological shapes are graphs \cite{I, II}. 
Moreover, identifying mirror images has no separate effect on the Claw graph.  
Thus 
\be
S_3 \times C_2 \m \mbox{ acts as } \m  D_3 \mbox{ (dihaedral group of order 6) } , 
\ee
permitting us to rewrite the labelling and mirror image based definition (\r{S(3, 2)-Top-Disc}) of the topological configuration in space as 
\be
\Top\mbox{-}\Leib_{\FrS}(3, 2)   \es   \frac{\Top\mbox{-}\FrS(3, 2)}{D_3}              \m . 
\ee
This is very natural since 
\be
Aut(\mbox{Claw}) =  D_3 \m , 
\ee
where $Aut$ denotes automorphism group.  

\m 

\n{\bf Structure 1} For arbitrary $(N, d)$, we denote our lattice of discrete quotients by $\lattice$, and its general member by $\Gamma^{\tau}$, according to   
\be 
\Top\FrS(N, d; \Gamma^{\tau}) \:=  \frac{\Top\FrS(N, d)}{\Gamma}  \mma  
\Gamma^{\tau} \, \in \, \lattice  \left( \m S_N \times C_2 \mbox{ distinct subgroup actions on the } \Top\FrS(N, d) \mbox{ graph }  \right)  \m . 
\ee

\subsection{Consequent lattice of 8 shape spaces}

\n For metric-level shapes, there are 8 cases. 

\m  

\n{\bf Case 0 = S$^{\prime}$)} $\FrS(3, 2; S_3 \times C_2)$ is the discrete quotient of the triangleland shape sphere under indentification of all three particles labels and of mirror images.

\m 

\n{\bf Case S)} $\FrS(3, 2; S_3)$ is the discrete quotient of the triangleland shape sphere under identification of all three particle labels.

\m

\n{\bf Case E$^{\prime}$)} $\FrS(3, 2; V_4)$ is the discrete quotient of the triangleland shape sphere by identification of 2 of the 3 particle labels and mirror image identification.

\m  

\n{\bf Case A$^{\prime}$)} $\FrS(3, 2; C_6)$ is the discrete quotient of the triangleland shape sphere by even permutations and mirror image identifications.

\m

\n{\bf Case E)} $\FrS(3, 2; C_2\mbox{-}\ml\ma\mb\me\ml)$ is the discrete quotient of the triangleland shape sphere 
                                                                         by just the identification of 2 of the 3 particle labels. 

\m

\n{\bf Case A)} $\FrS(3, 2; A_3) = \FrS(3, 2; C_3)$ is the discrete quotient of the triangleland shape sphere by just the even permutations.  

\m

\n{\bf Case 1$^{\prime}$)} $\FrS(3, 2; C_2\mbox{-}\mr\me\mf)$ is the discrete quotient of the triangleland shape sphere by just mirror image identification.

\m

\n{\bf Case 1)} $\FrS(3, 2; id)$ is just the shape sphere itself: 
\be 
\FrS(3, 2; id) = \FrS(3, 2) = \mathbb{S}^2  \m .
\ee   
\n{\bf Structure 1} We denote our lattice of discrete quotients by  $\lattice$, and its general member by $\Gamma$ according to   
\be 
\FrS(N, d; \Gamma) \:=  \frac{\FrS(N, d)}{\Gamma} \mma \Gamma \, \in \, \lattice  \left(  S_3 \times C_2 \mbox{ distinct subgroup actions on } \FrS(3, 2)  \right) \m .
\ee
This is depicted in Fig \r{S(3, 2)-Top-8}. 

\m

\n{\bf Remark 1} $\lattice$'s top element is the shape sphere $\FrS(N, d)$: the shape space with {\sl no} discrete quotienting.   

\m

\n{\bf Remark 2} $\lattice$'s bottom element is the {\it Leibniz space}  
\be 
\Leib_{\sFrS}(N, d) := \FrS(3, 2; S_3 \times C_2) \m : 
\ee
the shape space with the maximal amount of discrete quotienting. 
As the bottom element, it remains acceptable for it to have further special notation and nomenclature: Leibniz space and fundamental cell.  
Its being the bottom element justifies it having further names, yet further such being the {\it fundamental cell} of the shape space, and, 
                                                     in the specific case of triangleland, {\it Kendall's spherical blackboard}.  
 
\m 

\n{\bf Remark 3} For unsaturated spaces, 
$\FrS(N, d; C_2\mbox{-ref})$ has previously been known as $\FrS(N, d)$ in accounts which refer to {\sl our} $\FrS(N, d)$ as $\w{\FrS}(N, d)$. 

\m 

\n{\bf Remark 4} For unsaturated spaces, 
$\FrS(N, d; S_N)$ has previously been known by the less extendible notation of $\FrI\FrS(N, d)$, standing for (fully) indistinguishable-particle shape space. 

\m 

\n{\bf Remark 5} Our case notation for $N = 3$ is A) for $A_3$ involvement, 
                                                  E) for signature $\epsilon$ involvement, 
												  S) for $S_3$ involvement and 
												  $\mbox{}^{\prime}$ for mirror image identification. 
1) Is additionally a top lattice element notation, by which S$^{\prime}$) is cast as 0): the corresponding bottom element notation.  

\m 

\n{\bf Proposition 1} Aside from Case 1)'s already encountered hemisphere, we need to provide the topology of each of the above-defined shape spaces; 
we do so in a convenient order of geometrical development.  
%
{            \begin{figure}[!ht]
\centering
\includegraphics[width=0.65\textwidth]{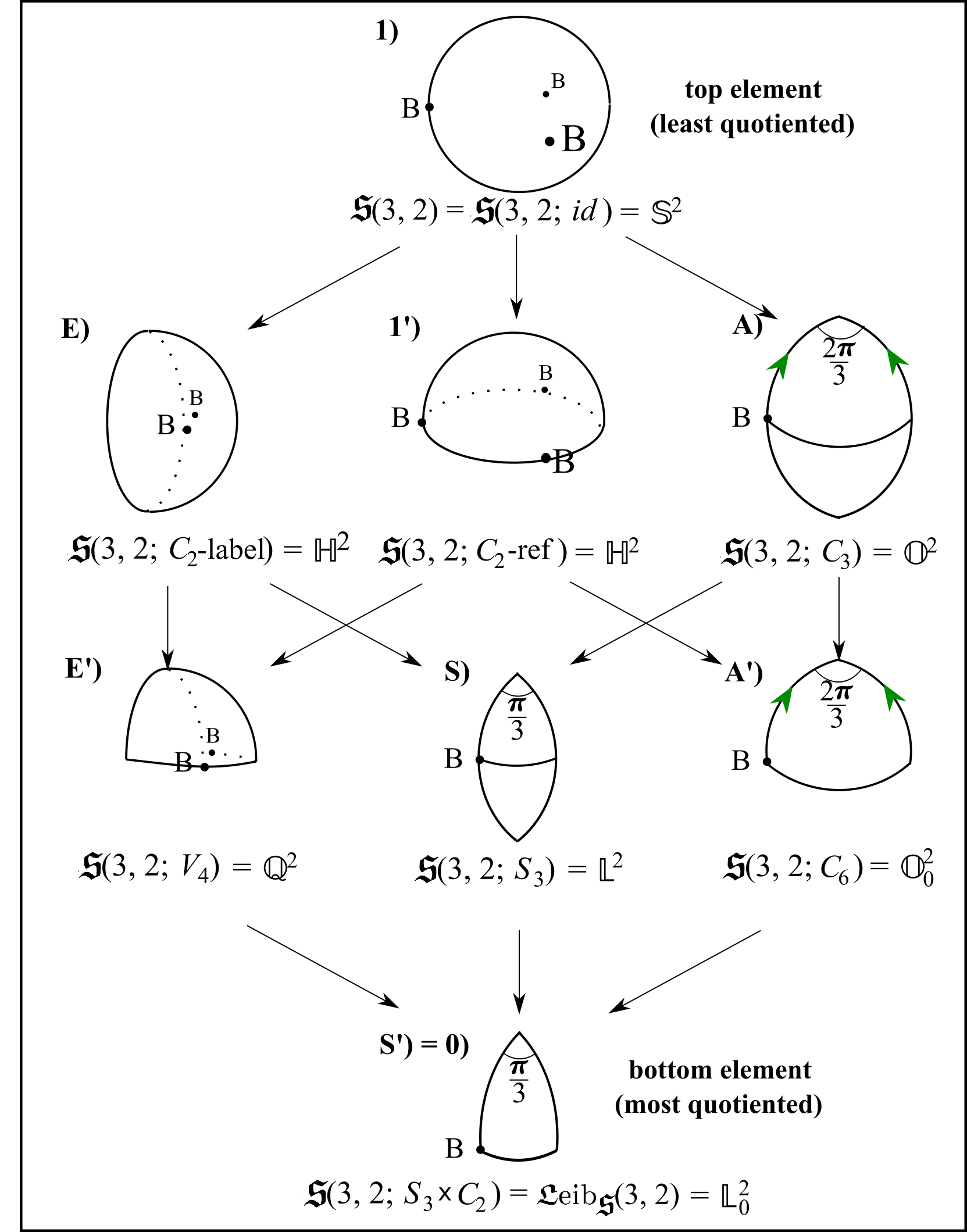}
\caption[Text der im Bilderverzeichnis auftaucht]{        \footnotesize{Antitone dual lattice of 8 shape spaces, using Appendix C's notation for each manifold-or-orbifold chunk.    
These last two's green arrows indicate topological identification; they are then redrawn in Fig \r{Orbifold-Up} with this topological identification made.} }
\l{S(3, 2)-Top-8} \end{figure}          }
%
{            \begin{figure}[!ht]
\centering
\includegraphics[width=0.3\textwidth]{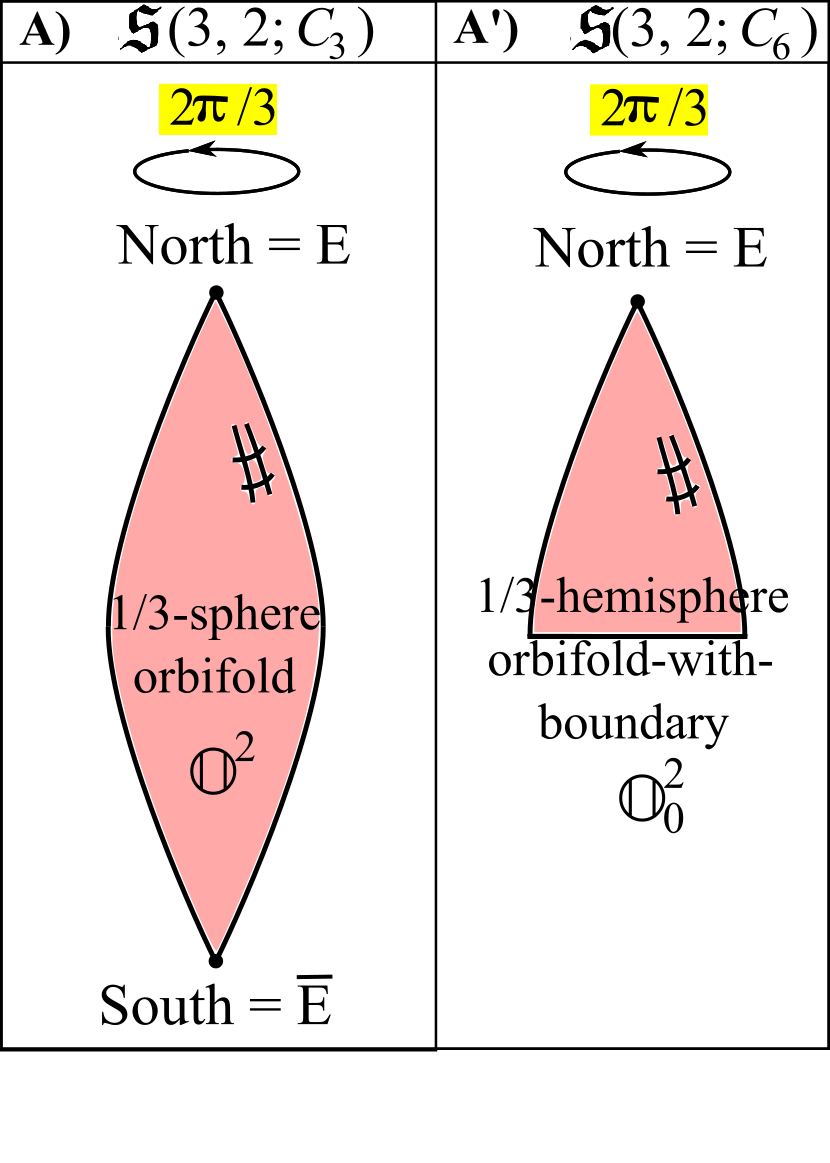}
\caption[Text der im Bilderverzeichnis auftaucht]{        \footnotesize{Folded version of the four orbifold constructs. 
Subsequent figures use the folded versions.
At the metric level, we identify A)'s two conical singularities, and A$^{\prime}$)'s single conical singularity, as lying at the equilateral triangle poles.} }
\l{Orbifold-Up} \end{figure}          }

\m 

\n{\bf Remark 5} With orbifolds starting to enter the current paper in Figs \r{S(3, 2)-Top-8}-\r{Orbifold-Up}, we point to \c{TF17} for a review of orbifolds, 
for all that these authors' applications are very different from ours.

\m

\n{\bf Remark 6} See Fig \r{S(3, 2)-Met-8} for the further metric-level decor on these shape spaces, i.e. where each type of triangle lies thereupon.  
%
{            \begin{figure}[!ht]
\centering
\includegraphics[width=0.7\textwidth]{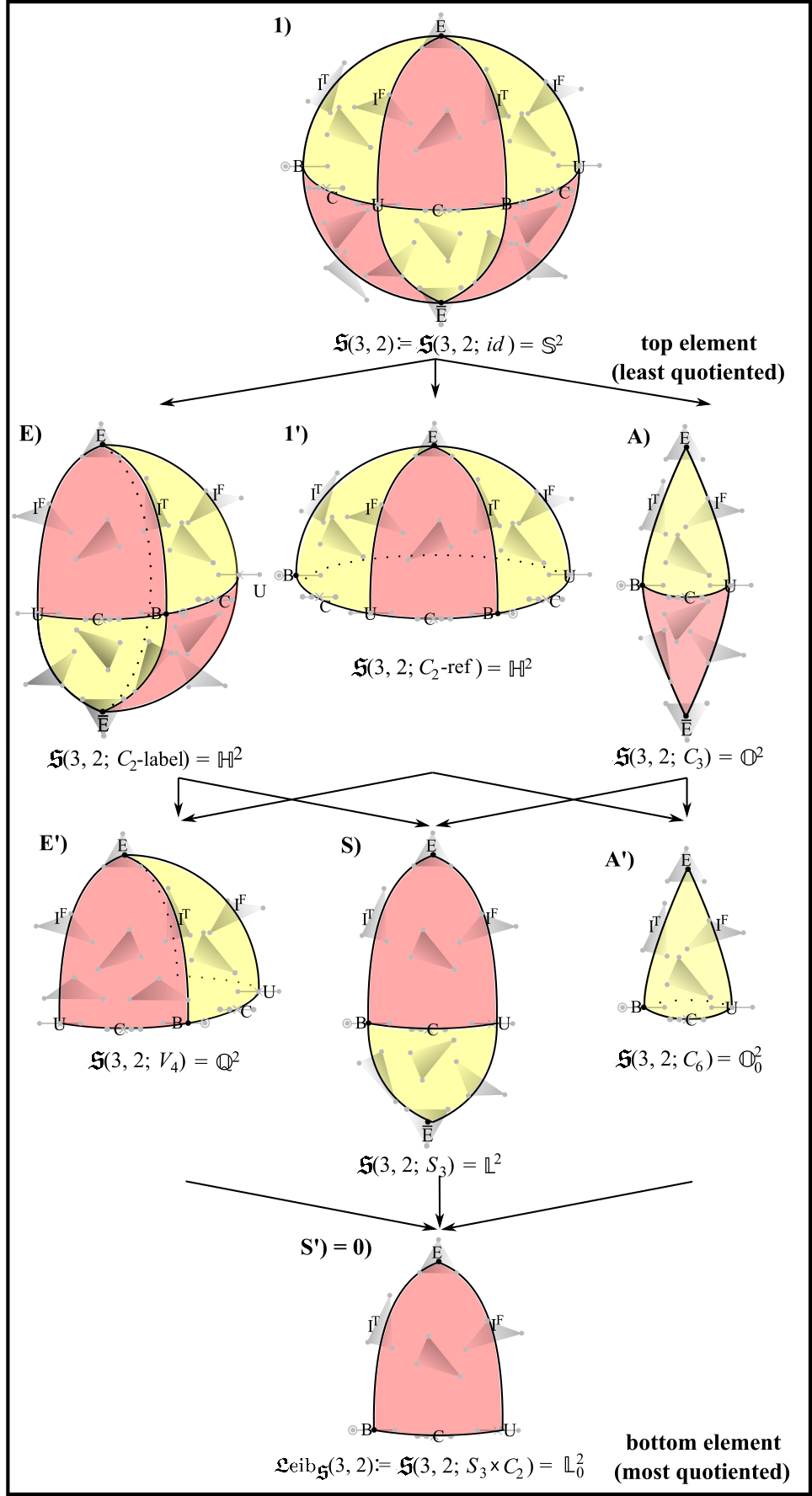}
\caption[Text der im Bilderverzeichnis auftaucht]{        \footnotesize{Antitone dual lattice of 8 shape spaces, now with metric-level decor indicated: 
which triangles in space correspond to which points on the triangleland shape sphere.
Each tessellation by Leibniz spaces is indicated with alternating pale yellow-and-red tiles.  } }
\l{S(3, 2)-Met-8} \end{figure}          }

\m

\n{\bf Remark 7} 1) and 1$^\prime$) correspond to $\tau$1), E) and E$^{\prime}$) to $\tau$2), and all of 0), 0$^{\prime}$), A) and A$^{\prime}$) to $\tau$0).
This gives the topological shapes' coarse-graining of the metric shapes' classification of modelling types. 

\vspace{10in}

\subsection{Corresponding lattices of relational spaces}
%
{            \begin{figure}[!ht]
\centering
\includegraphics[width=0.63\textwidth]{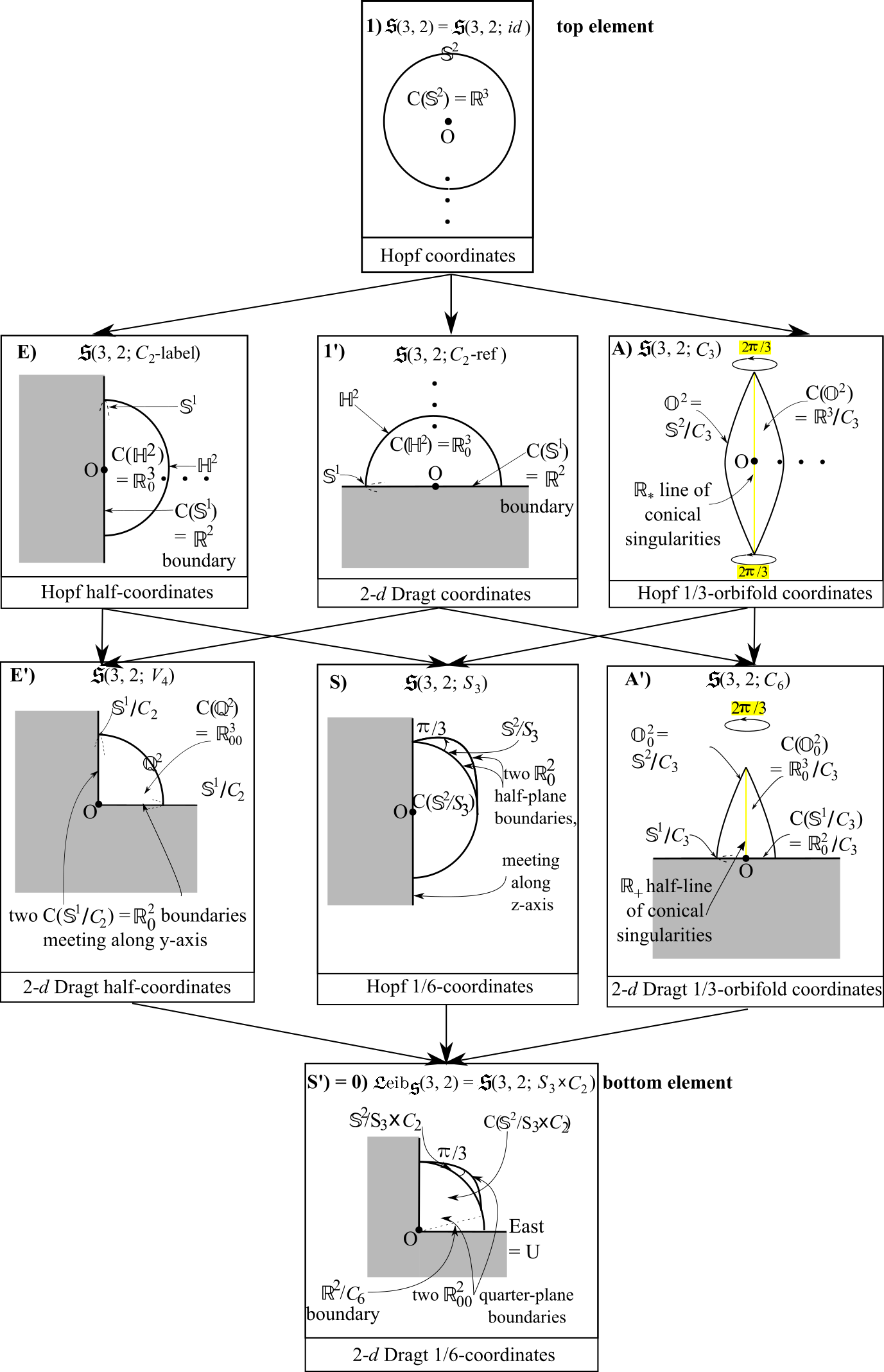}
\caption[Text der im Bilderverzeichnis auftaucht]{        \footnotesize{Lattice of relational spaces as required to understand monopole structure in the next section.
Orbifold structure is highlighted in bright yellow.
The coordinates mentioned under each case are (\r{Hopf-x}-\r{Hopf-z}) but with different coordinate ranges as indicated, 
with the 2-$d$ Dragt range being $area \geq 0$, i.e. corresponding to unsigned area.} }
\l{R(3, 2)-Met}\end{figure}            }
  
\n{\bf Remark 1} In each case, the corresponding relational spaces are topologically and metrically the cones over these shape spaces, 
with the maximal collision O in the role of cone point. 

\m 

\n{\bf Remark 2} For topological shapes, this gives 
\be 
\Top\mbox{-}{\cal R}(3, 2; \Gamma^{\tau}) = \mC(\Top\FrS(3, 2; \Gamma^{\tau}))  \mma  
\Gamma^{\tau} \, \in \, \lattice  \left( \m S_3 \times C_2 \mbox{ group actions on  Claw }  \m \right) \m .
\ee 
\n{\bf Remark 3} For metric-level shapes, this gives 
\be 
{\cal R}(3, 2; \Gamma) = \mC(\FrS(3, 2; \Gamma))                 \mma  
\Gamma \, \in \, \lattice  \left( \m S_3 \times C_2 \mbox{ group actions on } {\cal R}(3, 2) \m \right) \m , 
\ee 
this lattice set being isomorphic to 
\be 
\lattice  \left( \m S_3 \times C_2 \mbox{ group actions on } \FrS(3, 2) \m \right) \m . 
\ee 
\n{\bf Remark 4} See Fig \r{R(3, 2)-Met} for the lattice of triangleland relational spaces and associated Hopf-type coordinate ranges.  

\vspace{10in}

\section{Triangleland's suite of monopoles}\l{Monopoles}

\subsection{The eight cases}

\n{\bf Case 1)} On ${\cal R}(3, 2; id) = {\cal R}(3, 2)$, the distinguishable-particles mirror-images-distinct 2-$d$ 3-body problem's monopole 
is a configuration space realization of the {\it Dirac monopole} as depicted in Fig \r{Monopoles-8}.1).
The Dirac monopole is more usually realized in space instead.
We thus term the current example the {\it full} alias {\it double-cover triangleland realization of the Dirac monopole}.  

\m

\n{\bf Case 1$^{\prime}$)} On ${\cal R}(3, 2; C_2\mbox{-}\mr\me\mf)$, the distinguishable-particles mirror-images-identified 2-$d$ 3-body problem's monopole 
is the {\it Iwai-type half-space monopole} \cite{Iwai87} as depicted in Fig \r{Monopoles-8}.1$^{\prime}$).  

\m

\n{\bf Case E)} On ${\cal R}(3, 2; C_2\mbox{-}\ml\ma\mb\me\ml)$, the 2-indistinguishable-particles mirror-images-distinct 2-$d$ 3-body problem's monopole is another  
half-space monopole as depicted in Fig \r{Monopoles-8}.E).

\m 

\n{\bf Case E$^{\prime}$)} On ${\cal R}(3, 2; V_4)$, the 2-indistinguishable-particles mirror-images-distinct 2-$d$ 3-body problem's monopole is a 
$\frac{1}{4}$-{\it space monopole} as depicted in Fig \r{Monopoles-8}.E$^{\prime}$).

\m 

\n{\bf Case A)} On ${\cal R}(3, 2; C_3)$, the even-permutations-identified mirror-images-distinct 2-$d$ 3-body problem's monopole is of a new orbifold kind: 
the {\it $\frac{1}{3}$-space orbifold monopole}        as depicted in Fig \r{Monopoles-8}.A).

\m

\n{\bf Case A$^{\prime}$} On ${\cal R}(3, 2; C_6)$, the even-permutations-and-mirror-images-identified 2-$d$ 3-body problem's monopole 
                                                         is of a new orbifold-with-boundary kind: 
{\it $\frac{1}{3}$-half-space orbifold monopole}        as depicted in Fig \r{Monopoles-8}.A$^{\prime}$).

\m

\n{\bf Case S)} On ${\cal R}(3, 2; S_3)$, the indistinguishable-particles mirror-images-distinct 2-$d$ 3-body problem's monopole 
is the {\it 1/6th-segment monopole} as depicted in Fig \r{Monopoles-8}.S).

\m

\n{\bf Case $\bS^{\prime}$) = 0)} On ${\cal R}(3, 2; S_3 \times C_2) = \Leib_{{\cal R}}(3, 2)$, the indistinguishable-particles mirror-images-identified 2-$d$ 3-body problem's monopole 
																							is the {\it 1/6th-hemisegment monopole} as depicted in Fig \r{Monopoles-8}.0), 
																							alias {\it Leibniz--Kendall monopole} alias {\it fundamental cell monopole}

\subsection{Shape space charts corresponding to each}

\n{\bf Case 1)} Here `North = E' (pale blue\f{Figs \r{Monopoles-8} and \r{Monopoles-3-d-4} are a rare exception to our rule that pale blue denotes topological shapes 
and bright blue their spaces.}) and `South = $\overline{\mE}$' (pink) stereographic coordinate chart patches can be made large enough to overlap with each other.
For each chart, the Dirac string can be placed through the opposite pole so as to elude that chart.  

\m

\n{\bf Case 1$^{\prime}$)} Here a hemisphere $\mathbb{H}^2$'s worth of `North = orientationless equilateral triangle E' stereographic coordinate chart suffices to cover everything. 
The Dirac string can be placed through the physically unrealized `South = $\overline{\mE}$' hemisphere.  

\m

\n{\bf Case E)} Here a hemisphere  $\mathbb{H}^2$'s worth of `East = uniform collinear configuration U' (green) stereographic coordinate chart covers everything. 
So the Dirac string can be placed through the physically unrealized `West = binary collision B' hemisphere.  

\m

\n{\bf Case E$^{\prime}$)} Here a quadrant  $\mathbb{Q}^2$'s worth of either `North = E' or `East = U' stereographic coordinate chart suffices to cover this; 
on democratic grounds, we indicate this chart in cyan. 
So the Dirac string can be placed through the physically unrealized other three quadrants.  

\m 

\n{\bf Case S)} Here a $\frac{1}{6}$-lune $\mathbb{L}^2$'s worth of `East = U' stereographic coordinate chart covers everything
So the Dirac string can be placed through the physically unrealized other five segments.  
[This being the only non-hemisphere non-quadrant lune in the paper, we effectively drop a $\frac{1}{6}$ label on it.]

\m 

\n{\bf Case S$^{\prime}$) = 0)} Here a $\frac{1}{6}$-hemilune  $\mathbb{L}^2_0$'s worth of either `North = orientationless E' or `East = U' stereographic coordinate chart 
covers everything.
So the Dirac string can be placed through the physically unrealized other five-and-a-half segments.  

\m 

\n{\bf Case A)} Here one has `North = E' (blue) and `South = $\overline{\mE}$' (pink) $\frac{1}{3}$-sphere orbifold  $\mathbb{O}^2$ stereographic coordinate orbifold chart patches 
which are large enough to overlap with each other.
We indicate orbifold charts with {\sl bright} versions of our colouring scheme.  
For each orbifold chart, the Dirac string can be placed through the opposite pole so as to elude that orbifold chart.    

\m 

\n{\bf Case A$^{\prime}$)} Here a $\frac{1}{3}$-hemisphere orbifold with boundary $\mathbb{O}^2_0$'s worth of `North = orientationless E' stereographic coordinate orbifold chart 
suffices to cover everything. 
The Dirac string can be placed in the physically unrealized `South =  $\overline{\mE}$' $\frac{1}{3}$-hemisphere.                                  
%
{            \begin{figure}[!ht]
\centering
\includegraphics[width=0.72\textwidth]{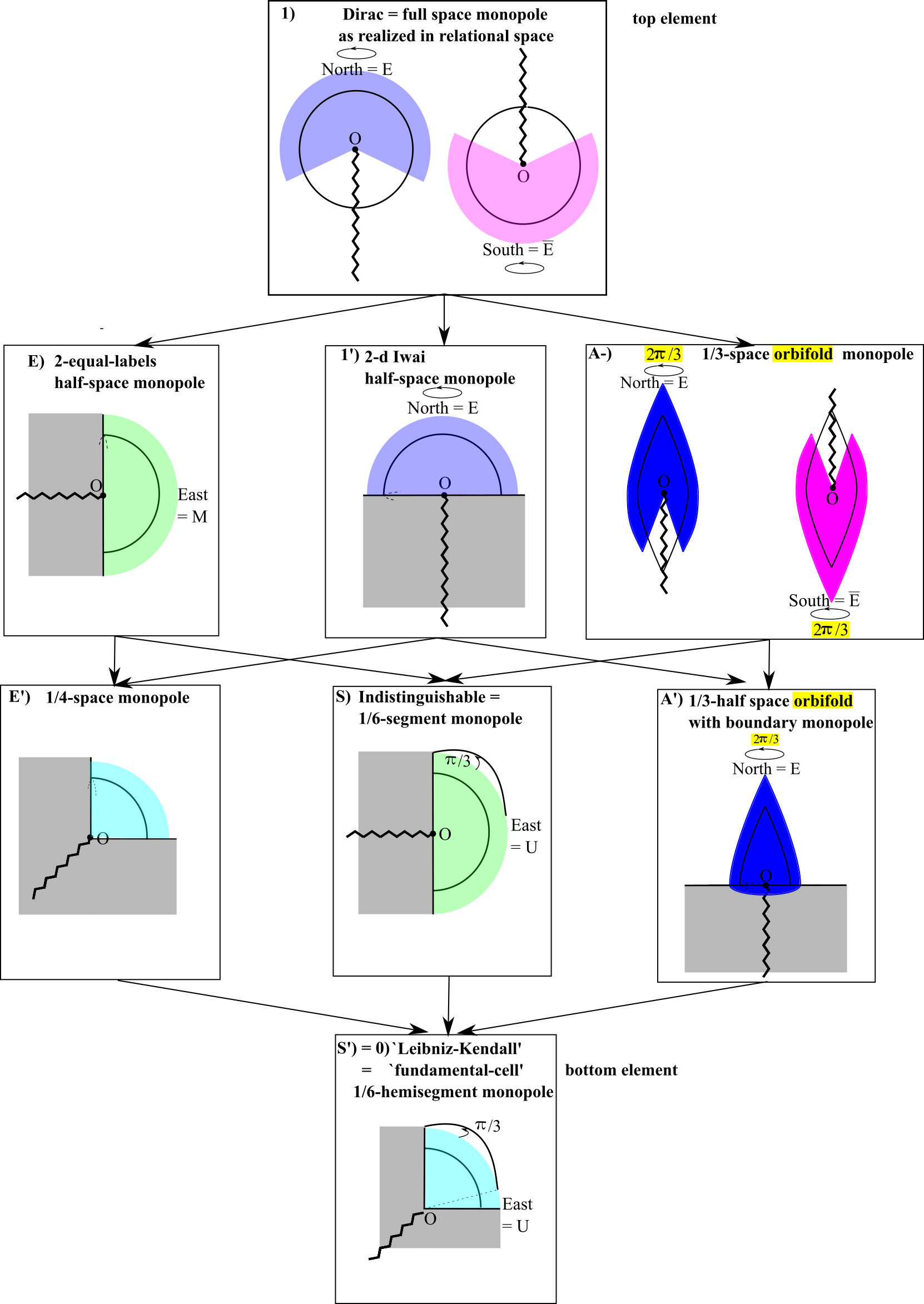}
\caption[Text der im Bilderverzeichnis auftaucht]{        \footnotesize{Relational triangleland suite of monopoles, with charts, 
                                                                                                                         Dirac string arrangements and 
																														 orbifold features marked.} }
\l{Monopoles-8} \end{figure}          }

\m

\n{\bf Remark 1} Counting up, one case needs two charts, and one needs two orbifold charts; these cases have nontrivial transition functions, 
whose phases induce topological quantization conditions, as we shall see in Appendix E. 
Five other cases can be described using a single chart, and the remaining case with a single orbifold chart; these do not elicit topological quantization conditions.  

\m

\n{\bf Remark 2} This account provides another reason to not conformally transform to the flat metric on relational space 
is that this would dispense with the monopole structure emanating from the maximal collision O (but this monopole encodes part of the physical properties of the system).

\subsection{Chern integral and total topological contributions: Gauss--Bonnet checked}
%
%
{            \begin{figure}[!ht]
\centering
\includegraphics[width=0.6\textwidth]{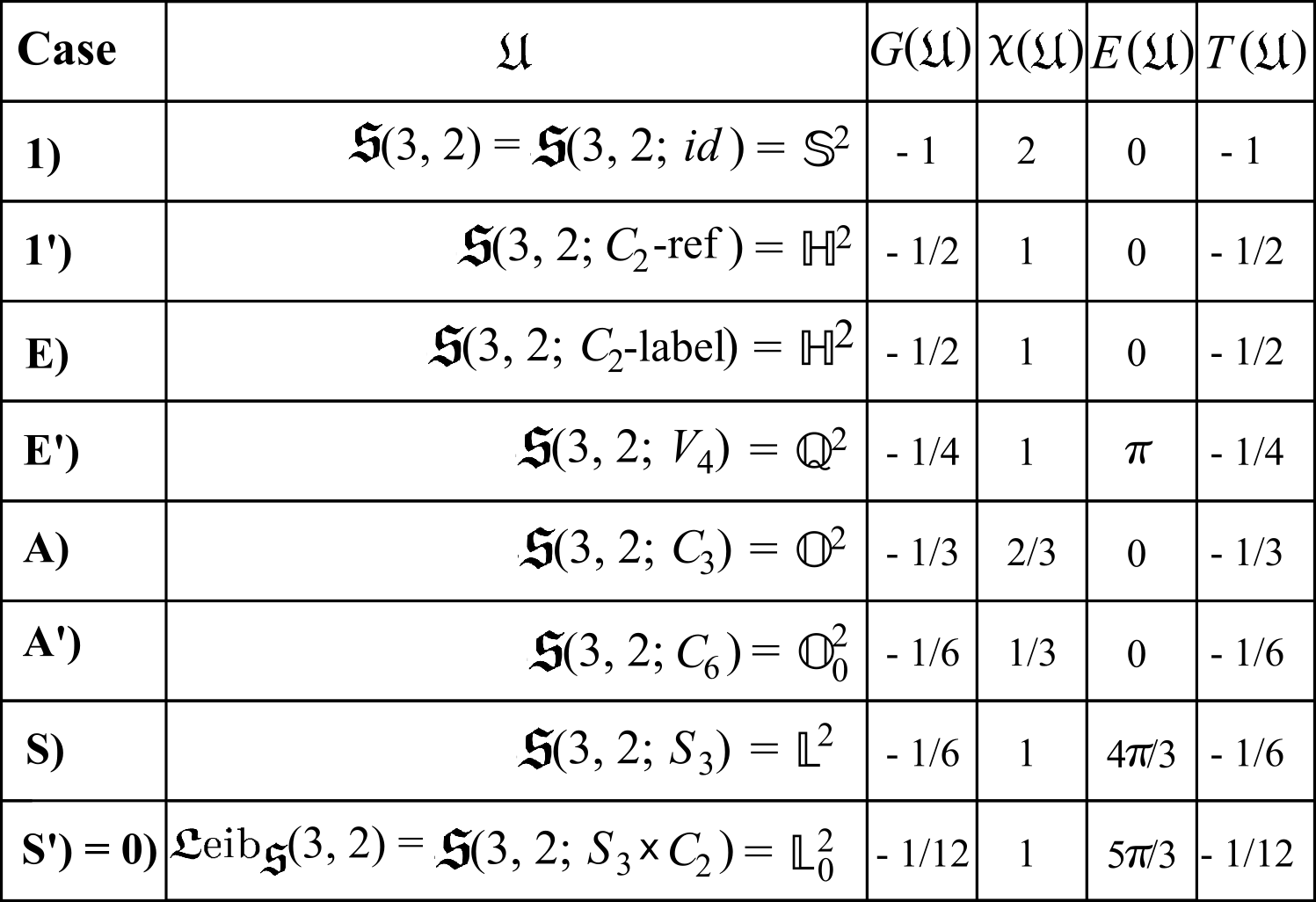}
\caption[Text der im Bilderverzeichnis auftaucht]{        \footnotesize{Using (\r{G(S)})                     for $G(\FrU))$, 
                                                                           Fig \r{Exterior-Angles}.a)'s counts for $\chi(\FrU)$  
                                                                       and Fig \r{Exterior-Angles}.b)'s for        $E(\FrU)$, 
																	   and    (\r{T(S)})                     for $T(\FrU)$,           
																	   we obtain $G(\FrU) = T(\FrU)$ concurrence in all cases: the Gauss--Bonnet--Chern Theorem holds.} }
\l{GB-Check} \end{figure}          }
%
{            \begin{figure}[!ht]
\centering
\includegraphics[width=0.8\textwidth]{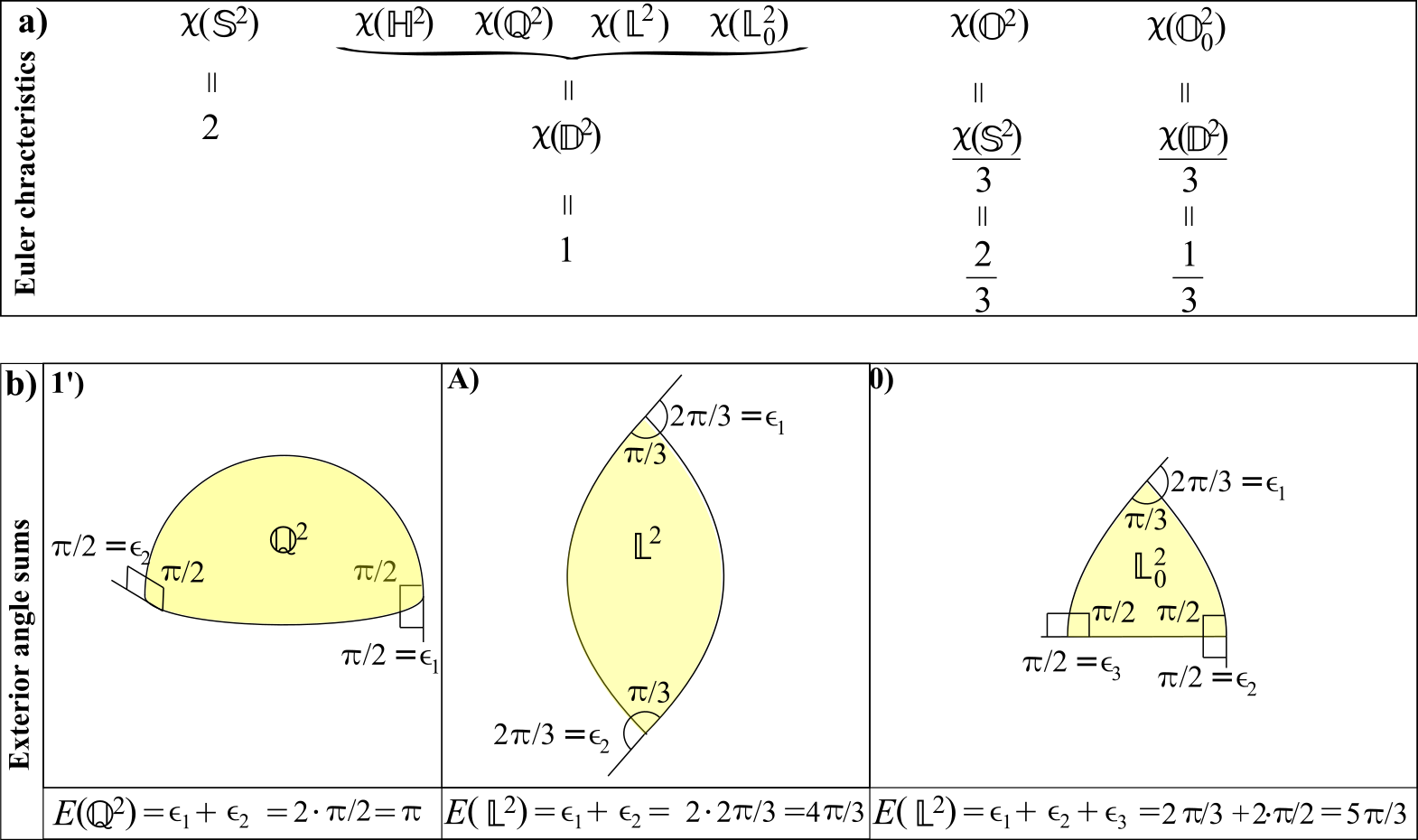}
\caption[Text der im Bilderverzeichnis auftaucht]{        \footnotesize{a) Euler characteristics for the 2-$d$ 3-body problem shape spaces. 
The last two of these use (\r{chi-orb}) with $m = 3$.
b) Exterior angles $\epsilon_I$ and their sums $E(\FrU)$ for cases E$^{\prime}$), S) and 0).
$E(\FrU) = 0$ for 1)          and A) by these having no boundary, 
              for 1$^\prime$)  and E) by their boundaries being complete great circles, 
		  and for A$^\prime$) by its boundary being a smoothly closed piece of the previous.} }
\l{Exterior-Angles} \end{figure}          }

\m 

\n{\bf Remark 1} These are given in Fig \ref{GB-Check} supported by the workings in Fig \ref{Exterior-Angles}.

\m

\n{\bf Remark 2} Topological distinction of shape spaces at metric level 
-- by combining the $\tau$ classification and Euler characteristic information -- provides a distinct coarse-graining of the metric-level cases.
This gives a 6-class `T' intermediary between the 3 tau classes and the 8 full metric level classes as per Fig \r{Class-Break}.
%
{            \begin{figure}[!ht]
\centering
\includegraphics[width=0.55\textwidth]{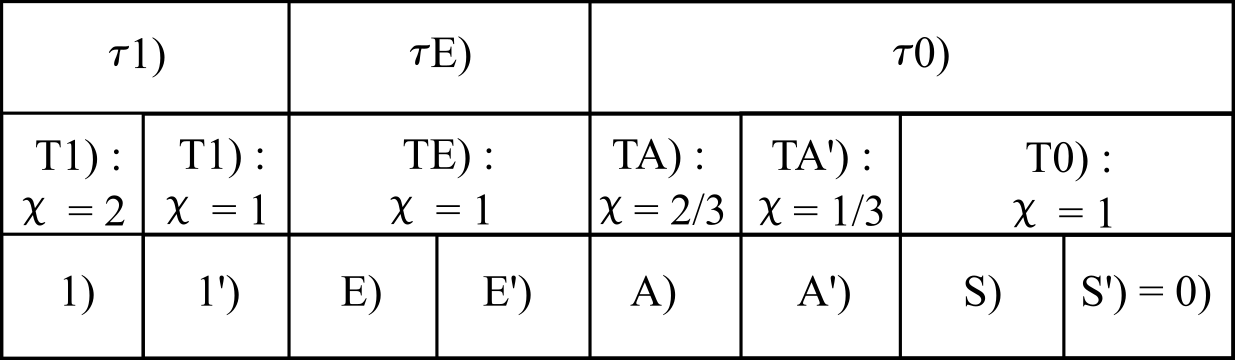}
\caption[Text der im Bilderverzeichnis auftaucht]{        \footnotesize{Fine-graining of the topological shape space classes $\tau$, 
         firstly  by the metric shapes' shape space topology's Euler characteristic $\chi$ giving the $T$ classes, 
and then secondly by the metric shapes' full metric shape spaces' classes.} }
\l{Class-Break} \end{figure}          }

\vspace{10in}

\section{Shape spaces and relational spaces for triangles in 3-$d$}

\n See the first part of Appendix B for the underpinning group theory.

\subsection{Topological shapes and shape spaces}

Our results concerning these are dimension-independent (for dimension $\geq 1$), as explained in \c{I, II, III}),  
so Secs \ref{TTS} and \ref{3-Chain}'s results carry over to the $d$ = 3 case verbatim.

\subsection{Corresponding shape space and relational space}

\n{\bf Structure IV$^{\prime}$} The shape space for this is 
\be 
\FrS(3, 3) \:=  \frac{\FrQ(3, 3)}{Sim(3)} 
           \es  \frac{\Frr(3, 3)}{Rot(3) \times Dil} 
		   \es  \frac{\FrP(3, 3)}{Rot(3)}  
           \es  \frac{\mathbb{S}^5}{SO(3)}		      \m . 
\ee 
This space turns out to be 
\be 
\FrS(3, 3) = \mathbb{H}^2_+ \, \coprod \, \mathbb{S}^1 \mbox{ } 
\ee 
with standard spherical metric (\r{Sphe-Met}), now with $\Theta$ taking the half-range $[0, \frac{\pi}{2}]$. 
$\mathbb{H}^2_+$ and $\mathbb{S}^1$ are strata and moreover fit together with trivial contiguity (in the sense of Appendix F).  
This bears some parallels with the mirror image identified case in 2-$d$. 
Now however only an $SO(2)$ subgroup of $SO(3)$ acts on the collinear configurations C, 
by which these constitute a distinct stratum (a disjoint orbit in the orbit space conceptualization).  

\mbox{ }

\n{\bf Structure V$^{\prime}$} The relational space for this is  
\be 
{\cal R}(3, 3) \:=  \frac{\FrQ(N, d)}{Eucl(3)} 
               \es  \frac{\Frr(N, d)}{Rot(3)}
			   \es  \frac{\mathbb{R}^6}{SO(3)}
			   \es \mC  \left( \mathbb{H}^2_+ \, \coprod \, \mathbb{S}^1  \right) 
			    =  \mathbb{R}^3_+ \, \coprod \, \mathbb{R}^2 \, \coprod \, \mO      \m . 
\ee
This has 2 strata (aside from the cone point: the maximal collision O) for the same reason as above. 
The $\mathbb{R}^3_+$ portion is moreover not metrically flat (albeit it is conformally flat if treated to the exclusion O, all as per Sec \r{Met-Config}).  
This case exhibits trivial contiguity as well.

\subsection{Corresponding lattice of shape spaces}
%
{            \begin{figure}[!ht]
\centering
\includegraphics[width=0.8\textwidth]{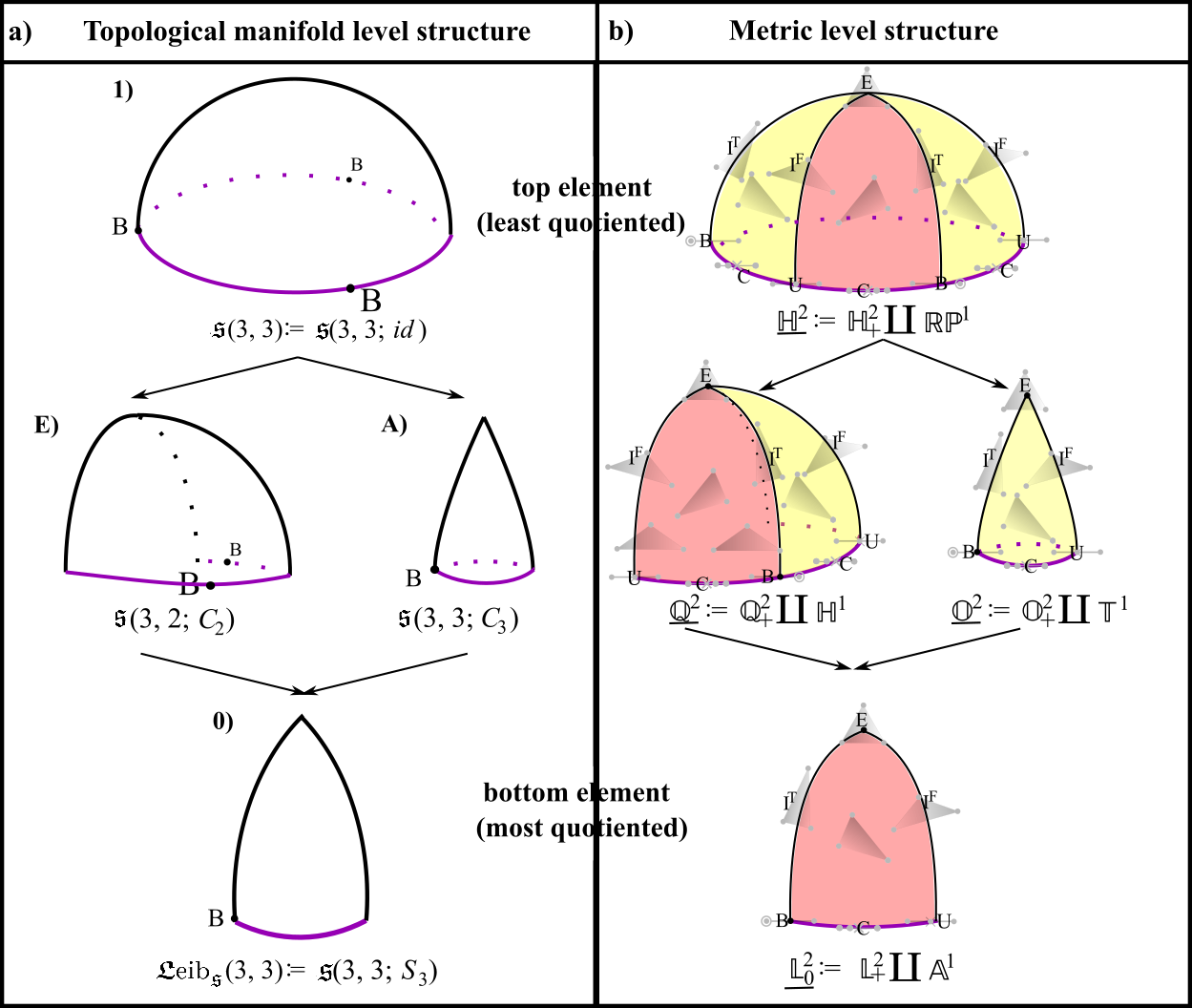}
\caption[Text der im Bilderverzeichnis auftaucht]{        \footnotesize{Lattice of 4 3-$d$ shape spaces of triangles at a) the topological and b) the metric geometry level,
with non-principal strata marked in purple. See Appendix C for the notation in use in the right half of the figure.} }
\l{S(3, 3)-Met} \end{figure}          }
%

$N = 3$ for $d = 3$ is an example of linear independence saturated model, necessitating a distinct shape space notation as now $S_3$ alone can act.  
Our choice of notation for this is to use lower-case $\Frs$ for the top shape space (discretely unquotiented),  
\be 
\Frs(N, d)  \:=  \frac{\FrS(N, d)}{C_2} \m : \m \mbox{ the top element for $d \geq N$ usually denoted } \m   \FrS(N, d)  \m , 
\ee
and then 
\be 
\Frs(N, d; \Delta) \:=  \frac{\Frs(N, d)}{\Delta}
\ee
for 
\be 
\Delta \in \lattice\left( \mbox{  subgroups of $S_N$ that act distinctly on $\FrS(N, d)$ } \right )  \m. 
\ee   
With this notation in mind in the case $N = 3, d = 3$, the below text and Fig \r{S(3, 3)-Met} provide the 4 possible discrete quotients for this: 
names, notation, topology, geometry and shapes-in-space to shape space map.

\m 

\n{\bf Definition 1} $\Frs(3, 3; S_3)$ is the discrete quotient of the triangleland shape hemisphere-with-edge under indentification of all three particles labels. 

\m 

\n{\bf Remark 1} We use a notation that is shared with the unsaturated case for the lattice's bottom element, 
\be 
\Leib_{\sFrS}(N, d) := \Frs(N, d; S_N)  \m \mbox{ for $d \geq N$}  \m .
\ee
This reflects that, at the bootom end, {\sl obligatory} mirror image identification plus full label identification in the saturated case 
coincides with the maximal quotient by the {\sl optional} mirror image identification plus full label identification in the unsaturated case.  

\m 

\n{\bf Definition $\u{\bA}$} $\Frs(3, 3; A_3) = \FrS(3, 3; C_3)$ is the discrete quotient of the triangleland shape hemisphere-with-edge  by just the even permutations.  

\m

\n{\bf Definition $\u{\bE}$} $\Frs(3, 3; C_2\mbox{-}\ml\ma\mb\me\ml)$ is the discrete quotient of the 3-$d$ triangleland shape hemisphere-with-edge  
                                                                         by just the identification of 2 of the 3 particle labels. 

\m

\n{\bf Case $\u{\bf 1}$} $\FrS(3, 3; id)$ is another realization of the shape hemisphere,  
\be 
\Frs(3, 3; id)  =  \Frs(3, 3) 
                =  \underline{\mathbb{H}}^2  \m :  
\ee   
the {\it hemisphere with edge constituting a separate stratum}.
This is stratified according to 
\be 
\u{\mathbb{H}^2} := \mathbb{H}^2_+ \, \coprod \mathbb{S}^1 \m, 
\ee 
where   the first -- and principal --  stratum is of generic triangular configurations                on which $SO(3)$ acts fully, 
whereas the second                     stratum is of collinear (including binary collision) triangles on which just an $SO(2)$ subgroup acts.
The latter is topologically $\mathbb{S}^1$ but has the metric half-angle range of $\mathbb{RP}^1$. 
In our `northern hemisphere' convention for mirror image identified triangles, the edge stratum lies horizontally underneath the principal stratum.  
This accounts for our choice of underline notation for this stratified space, and likewise for the seven further uses of the underline notation below.  

\m 

\n{\bf Proposition 1} The other 3 modelling possibilities' shape spaces have the following topological and stratificational structure.  

\m

\n{\bf Case $\u{\bE}$)}  
\be 
\sFrS(3, 3; C_2) = \underline{\mathbb{Q}^2} := \mathbb{Q}^2_+ \, \coprod \, \mathbb{Q}^1 \m, 
\ee 
the {\it quadrant with bottom edge constituting a separate stratum}, as depicted in Fig \r{S(3, 2)-Top-8}.3$^{\prime}$).  

\m   

\n{\bf Case $\u{\bA}$)}  
\be
\sFrS(3, 2; C_3) = \underline{\mathbb{O}_0^2} := \mathbb{O}_+^2 \, \coprod \, \mathbb{O}^1 \m, 
\ee 
the $\frac{1}{3}${\it-hemisphere orbifold with edge constituting a separate stratum} (Fig \r{S(3, 2)-Top-8}.2$^{\prime}$).
[This being the only orbifold in this paper, we effectively drop a $\frac{1}{6}$ label on it.]   
The principal stratum $\mathbb{O}_+^2$ is the open version of the $\frac{1}{3}$-orbifold with edge, still possessing a single conical singularity at the equilateral triangle E. 
The lower stratum $\mathbb{T}^1$ is topologically $\mathbb{S}^1$ but has the metric $\frac{1}{3}$-angle range. 
It is the still-periodic topological $\frac{1}{3}$-circle counterpart to $\mathbb{RP}^2$'s half-circle. 
In 2-$d$, the still-periodic $\frac{1}{3}$-sphere is an orbifold, which we denote by $\mathbb{O}^2$. 
While $\mathbb{T}^1$ is partly analogous to this, it is not itself an orbifold out of having no conical singularity. 
Thus we use a new notation $\mathbb{T}$, out of reserving $\mathbb{O}$ to clearly pick out spaces with orbifold features. 

\m 

\n{\bf Case $\u{\bf 0}$)} 
\be 
\Leib_{\tFrS}(3, 2) := \FrS(3, 2; S_3) = \underline{\mathbb{L}_0^2} := \mathbb{L}_+^2 \, \coprod \, \mathbb{L}^1 \m,
\ee 
the $\frac{1}{6}$-{\it hemilune with bottom edge constituting a separate stratum} (Fig \r{S(3, 2)-Top-8}.4$^{\prime}$).  
The principal stratum $\mathbb{O}_+^2$ is the version of the $\frac{1}{6}$-hemilune with edge with closed side-edges and open bottom edge 
(the bottom two corners having open status as well). 
The lower stratum $\mathbb{A}^1$ is the $\frac{1}{6}$-arc with closed endpoints; this being the only non-quadrant arc stratum, we effectively drop a $\frac{1}{6}$ label on it. 

\m

\n{\bf Remark 2} See Fig \r{S(3, 3)-Met}.b) for the further metric-level decor on these shape spaces, i.e. where each type of triangle lies thereupon.  

\m

\n{\bf Remark 3} $\u{1}$) corresponds to $\tau$1) and $\u{\mE}$) to $\tau$2), whereas $\u{\mA}$ and $\u{0}$ both correspond to $\tau$3).
In 3-$d$, the T classification turns out to be equally fine to the full classification rather than being a coarse-graining of it.  

\m

\n{\bf Remark 4}  While these cases in many ways correspond to 4 of the 2-$d$ cases according to the 'unprimed and primed' to 'underlined' correspondence, 
all four of the underlined cases have distinct-stratum pieces consisting of collinear shapes C, as compared to their most direct (primed) counterparts.   

\m 

\n{\bf Remark 5} Sec 7.3 and Appendix F contain further exposition about stratification in Shape Theory.

\subsection{Corresponding lattices of relational spaces}

\n{\bf Remark 1} In each case, the corresponding relational spaces are topologically and metrically the cones over these shape spaces, 
with the maximal collision O in the role of cone point. 
 
\m 

\n{\bf Remark 2} For topological shapes, this gives   
\be 
\Top\mbox{-}\scR(3, 3; \Delta^{\tau}) = C(\Top\mbox{-}{\scR}(3, 3; \Delta^{\tau})) \mma  
\Delta^{\tau} \, \in \, \lattice  \left(  S_3  \mbox{ distinct subgroup actions on C(Claw) }  \right)  \m .
\ee 

\m 

\n{\bf Remark 3} For metric-level shapes, this gives  
\be 
\scR(3, 3; \Delta) = \mC(\sFrS(3, 3; \Delta) \mma  
\Delta \, \in \, \lattice  \left(  S_3  \mbox{ distinct subgroup actions on } \sFrS(3, 3)  \right)  \m .
\ee 
\n{\bf Remark 4} See Fig \r{R(3, 3)-Met} for the lattice of trianglelnd relational spaces and associated Hopf-type coordinate ranges.    
%
{            \begin{figure}[!ht]
\centering
\includegraphics[width=0.5\textwidth]{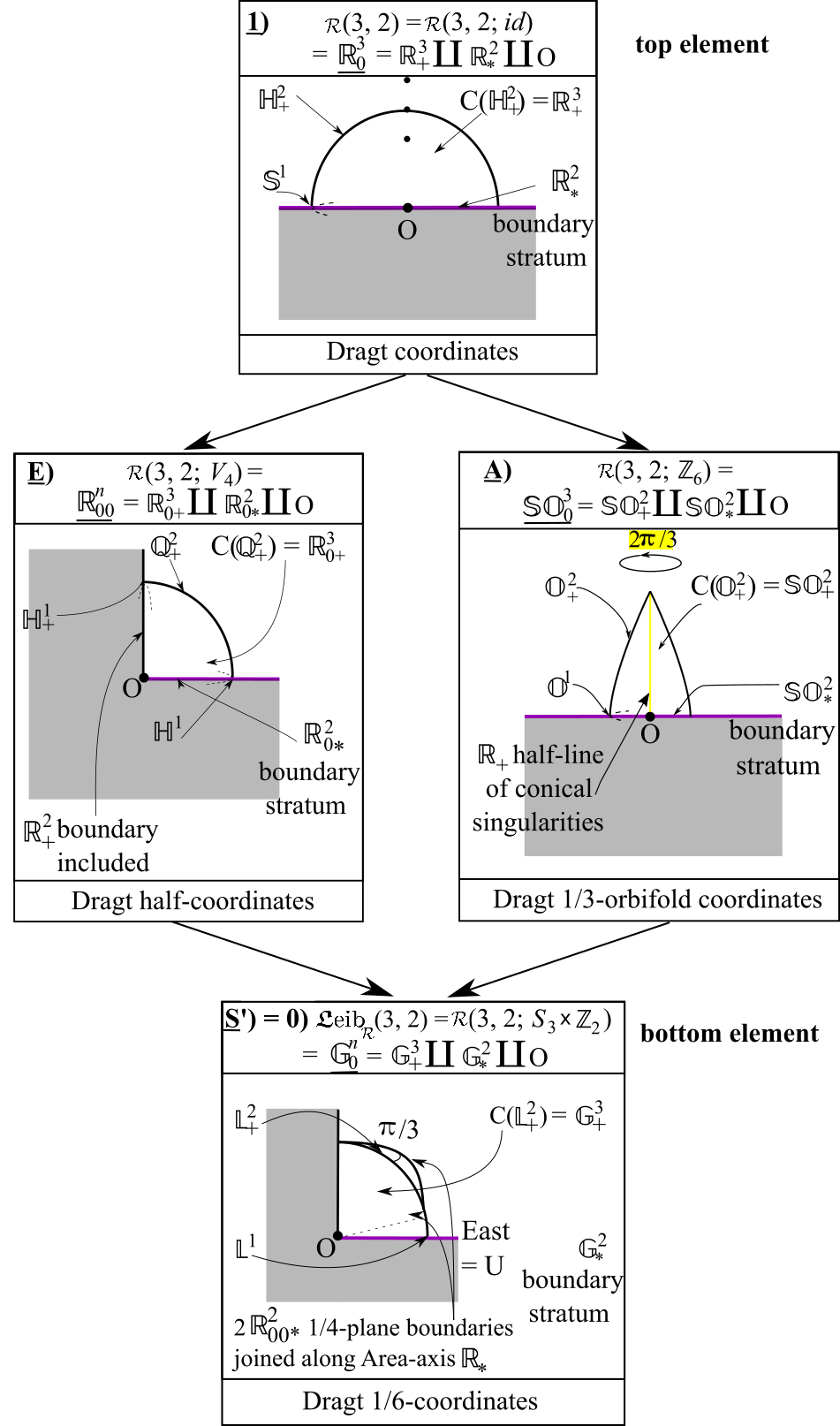}
\caption[Text der im Bilderverzeichnis auftaucht]{        \footnotesize{Lattice of 3-$d$ relational spaces; 
(implicitly 3-$d$) Dragt coordinates are (\r{Hopf-x}--\r{Hopf-z}) with range modified to $area > 0$.
} }
\l{R(3, 3)-Met}\end{figure}            }

\m 

\n{\bf Remark 5} Coning the boundary produces two strata: the intermediate stratum and the cone point as a separate bottom stratum.
This is captured case by case by Fig \r{Chunk}.  

\m

\n{\bf Remark 6} Thus we supplement the long-standing result that the relational space is the cone of shape space \cite{LR97, Cones, FileR} as follows for (3, 3) models. 

\m 

\n On the one hand, the relational space top stratum is the cone of the shape space top stratum. 

\m 

\n On the other hand, the cone of the shape space bottom stratum -- the edge of collinearities C -- 
splits into the relational space intermediate stratum with the cone point maximal collision O itself constituting a separate relational space bottom stratum.

\section{(3, 3) Monopoles}

\subsection{4 monopoles}

\n{\bf Case $\u{\bf 1})$} On 
\be 
\scR(3, 3; id)  =  \scR(3, 3) 
                =  \u{\mathbb{R}_0^3} 
			    =  \mC(\u{\mathbb{H}^2}) 
               \es	\mathbb{R}^3_{+} \, \coprod \, \mathbb{R}^2_{*} \, \coprod \, \mO		   \m ,
\ee 
the distinguishable-particles 3-$d$ 3-body problem's monopole is the {\it Iwai half-space monopole} \cite{Iwai87} as depicted in Fig \r{Monopoles-3-d-4}.$\u{1}$).  

\m 

\n{\bf Case $\u{\bE})$} On 
\be 
\scR(3, 3; C_2)  =  \u{\mathbb{R}_{00}^3} 
                 =  \mC(\u{\mathbb{Q}^2})
				\es \mathbb{R}^3_{0+} \, \coprod \, \mathbb{R}^2_{0*} \, \coprod \, \mO 
\ee 
the 2-indistinguishable-particles 3-$d$ 3-body problem's monopole is of a new kind: the {\it quadrant monopole}           as depicted in Fig \r{Monopoles-3-d-4}.$\u{\mA}$).

\m

\n{\bf Case $\u{\bA})$} On 
\be 
\scR(3, 3; C_3) = \u{\mathbb{SO}_0^3} 
                = \mC(\u{\mathbb{O}_0^2}) 
               \es \mathbb{SO}^3_{+} \, \coprod \, \mathbb{SO}^2_{*} \, \coprod \, \mO  				\m ,
\ee
the even-permutations-identified 3-$d$ 3-body problem's monopole is the {\it $\frac{1}{3}$-sphere orbifold monopole}        as depicted in Fig \r{Monopoles-3-d-4}.$\u{\bE}$).

\m

\n{\bf Case $\u{\bf 0})$} On 
\be 
\Leib_{\tcR}(3, 3) =  \scR(3, 2; S_3) 
                   =  \u{\mathbb{G}_0^3} 
				   =  \mC(\u{\mathbb{L_0^2}})
				  \es \mathbb{L_+^3} \, \coprod \, \mathbb{L^2} \, \coprod \, \mO                      \m, 
\ee 				   
the indistinguishable-particles 3-$d$ 3-body problem's monopole 
                                                                                           alias {\it Leibniz--Kendall monopole} alias {\it fundamental cell monopole} 
																						is the {\it 1/6th-hemilune monopole} as depicted in Fig \r{Monopoles-3-d-4}.$\u{0}$).

\subsection{Shape space charts corresponding to each}

\n{\bf Case $\u{\bf 1}$)} Here a hemisphere's worth of `North = orientationless equilateral triangle E' stereographic coordinate chart suffices to cover everything. 
The Dirac string can be placed through the physically unrealized `South = $\overline{\mE}$' hemisphere.  

\m

\n{\bf Case $\u{\bE}$)} Here a quadrant's worth of either `North = E' or `East = M' stereographic coordinate chart suffices to cover this; on democratic grounds, 
we indicate this chart in cyan. 
This permits the Dirac string to be placed through the physically unrealized other three quadrants.  

\m 

\n{\bf Case $\u{\bf 0}$)} Here a $\frac{1}{6}$-hemilune $\u{\mathbb{L_0}}$'s worth of either `North = orientationless E' or `East = M' stereographic coordinate chart 
covers everything.
This permits the Dirac string to be placed through the physically unrealized other five-and-a-half segments.  

\m 

\n{\bf Case $\u{\bA}$)} Here a $\frac{1}{3}$-hemisphere orbifold $\u{\mathbb{O_0}}$'s worth of `North = orientationless E' stereographic coordinate orbifold chart 
suffices to cover everything. 
The Dirac string can be placed through the physically unrealized `South =  $\overline{\mE}$' $\frac{1}{3}$-hemisphere.                                  
%
{            \begin{figure}[!ht]
\centering
\includegraphics[width=0.38\textwidth]{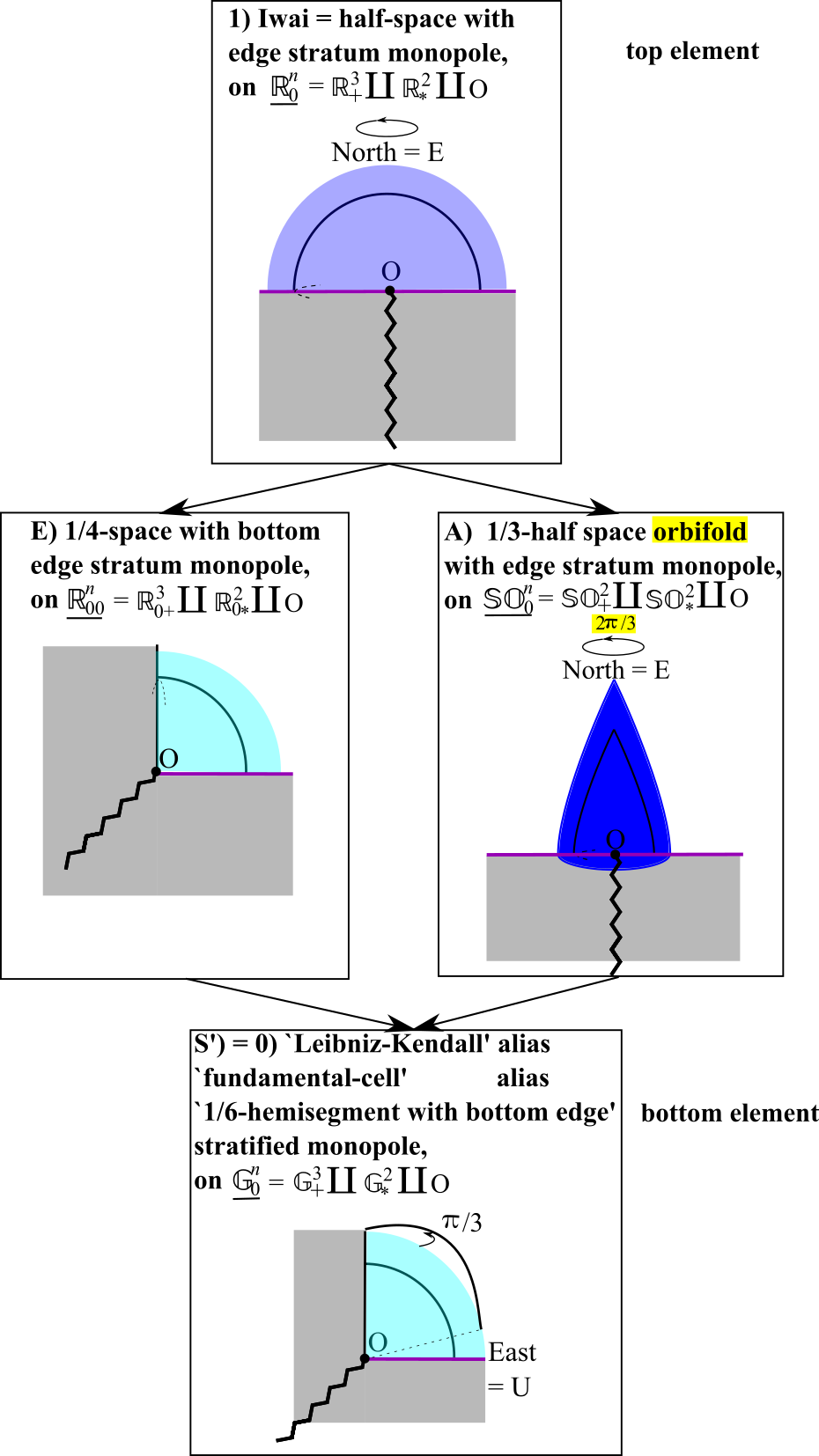}
\caption[Text der im Bilderverzeichnis auftaucht]{        \footnotesize{Relational triangleland monopoles 
in their most physically natural 3-$d$ space setting.   } }
\l{Monopoles-3-d-4} \end{figure}          }

\m

\n{\bf Remark 1} Counting up, three cases require a single chart, and the fourth requires a single orbifold chart; thus none of these elicit topological quantization conditions.  

\m

\n{\bf Remark 2} Following on from the previous section, another reason to not conformally transform to the flat metric on relational space 
is that this would dispense with the monopole structure emanating from O (but this is part of the physical properties of the system).   

\m

\n{\bf Remark 3} The Chern integrals work out as before 
[in the 4 2-$d$ cases 1$^{\prime}$), E$^{\prime}$), A$^{\prime}$) and 0) that most directly correspond to the four 3-$d$ cases]. 

\m 

\n{\bf Remark 4} So do the Euler characteristics and exterior angles, and consequently the total topological contributions are as before as well,   
so the Gauss--Bonnet Theorem checks carry through as before.  

\m 

\n{\bf Remark 5} Whether further subtleties arise from the monopoles' stratification features remains to be investigated.  
{\sl Physical} investigation of topological defects in the presence of stratification remains in its infancy \cite{ER90, RSV02}.  

\vspace{10in}

\section{Conclusion}

\subsection{Summary of results}

\n We consider some new shape spaces and relational spaces, which arise as discrete quotients under diverse modelling assumptions which combine (partial) particle label distinction, 
subtle differences between spatial dimension 2 and $\geq 3$, and, in the 2-$d$ case, the option of whether to identify mirror images.  
These modelling assumptions correspond to the distinctly-acting subgroups of $S_3 \times C_2$ in 2-$d$ and of $S_3$ in 3-$d$. 
3-$d$ 3-particle models' loss of the $C_2$ factor is due to mirror image identification now being obligatory by rotation through the third dimension. 
This diversity of subgroup actions constitutes a lattice, 
with the resulting quotient configuration spaces -- shape spaces and relational spaces -- forming antitone dual versions of this lattice.  
See the Summary Figure \r{Monopoles-Summary} for some key features of these shape spaces, including a layer-by-layer analysis 
involving coarse-grained lattices of topological shapes' shape spaces and of metric shapes' shape spaces' purely-topological features.  
For the lattices involved themselves, see Appendix B and Figures 8), 9) and 16).      

\m 

\n Each relational space is the cone over the corresponding shape space, corresponding to the inclusion of overall scale.  
We supplement this result \cite{LR97, FileR} as follows for 3-particle models in 3-$d$ with the following stratum-by-stratum decomposition of this coning.  

\m 

\n On the one hand, the relational space top stratum is the cone of the shape space top stratum. 

\m 

\n On the other hand, the cone of the shape space bottom stratum -- the edge of collinearities C -- 
splits into the relational space intermediate stratum with the cone point maximal collision O itself constituting a separate relational space bottom stratum. 

\m 

\n All in all, for (similarity and Euclidean) 3-body problems, the stratification is particularly well-behaved, relative to the other examples given in Appendix F.   
	
\m 

\n Each of the above two lattices of relational spaces resulting from the diversity of modelling assumptions produces an isomorphic lattice of distinct 
3-body problem monopoles realized in relational space.  
Six of these monopoles are new to the current paper, whereas four others were recently introduced in \cite{III}; the other two originate in the work of Iwai \cite{Iwai87} and, 
in an initially distinct setting -- in space, rather than configuration space -- Dirac \cite{DirMon, Dirac48}.  
We summarize their key properties in Fig \r{Monopoles-Summary}. 
%
{            \begin{figure}[!ht]
\centering
\includegraphics[width=1.0\textwidth]{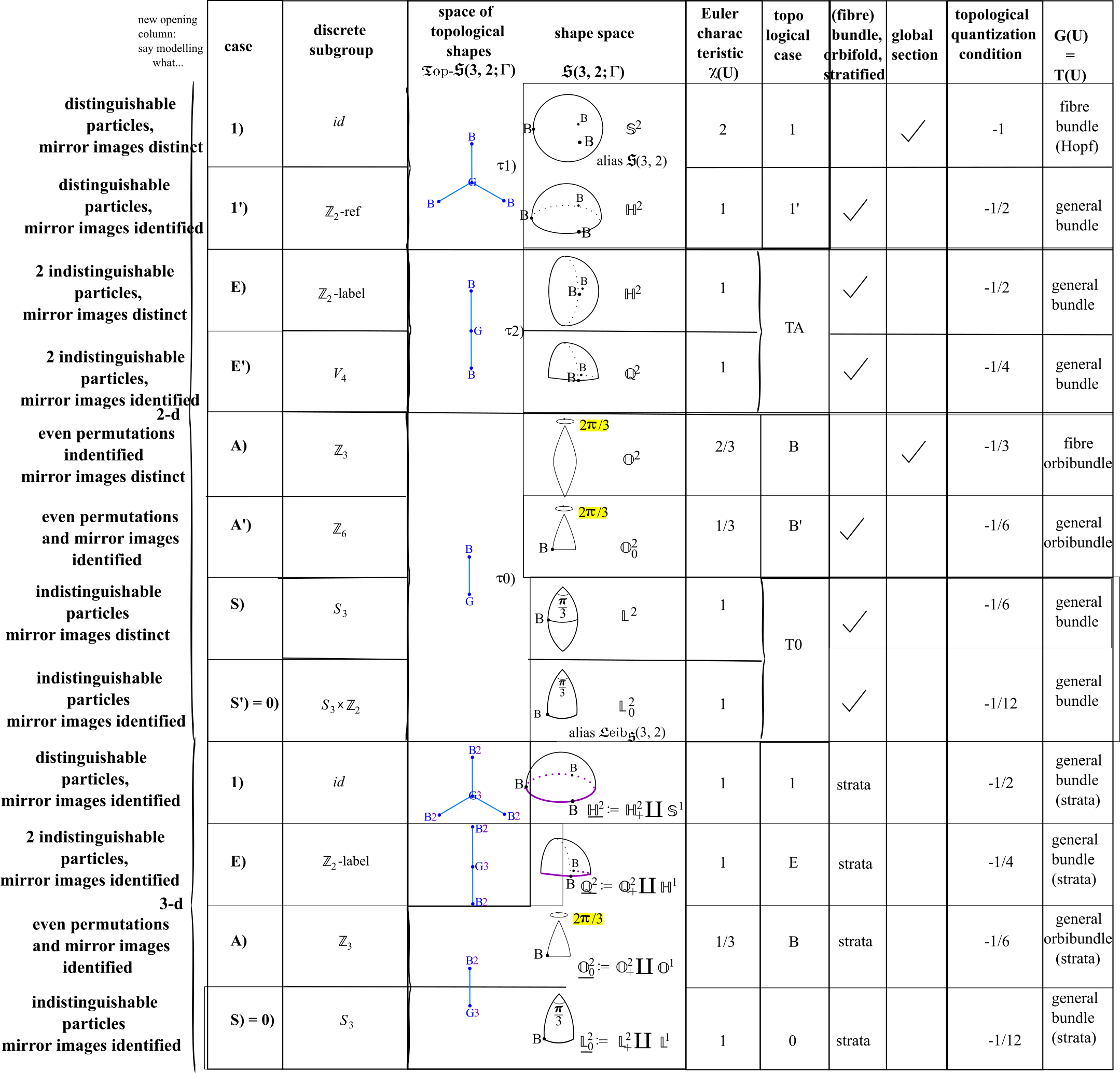}
\caption[Text der im Bilderverzeichnis auftaucht]{        \footnotesize{Summary table of the paper's twelvefold of modelling assumptions, 
corresponding distinctly-acting subgroups, shape spaces, relational spaces and monopoles with principal features as tabulated.  
The two cases with no global section are each covered by a pair of local sections, plain in the first case and orbifolded in the second. 
It is each pair's transition function which provides the consequently complementary topological quantization condition. 
To include that there is one transition function between local sections.
This figure is to be supplemented with relational space structure as explained in Figs 11 and 17. } }
\l{Monopoles-Summary} \end{figure}          }
	
\m	

\n The above multiple uses of lattices is part of a growing trend in using Order and Lattice Theory more widely in Foundational Theoretical Physics, as per Appendix A.

\subsection{Extension 1: $N$-a-gonland monopoles}

\n It is clear that this paper's work has a tractable 2-$d$ $N$-body problem $N$-a-gon configurations counterpart \cite{IV}. 
This is based upon the mathematics of the generalized Hopf map of Fig \r{Generalized-Hopf-Map}, with subgroup, quotienting and occasional orbifold considerations following suit.

\subsection{Extension 2: quantization of shapes}

The current paper's topological configurations underpin how one is to quantize the corresponding variety of 3-body problems \cite{QLS, Quantum-Triangles}.
This is clear from quantization's topological sensitivity as expounded in \cite{I84}.
These quantum models are moreover of foundational value, not only as quantizations but as background-independent and quantum-cosmological models as well 
\cite{FileR, APoT2, ABook, Project-1, Project-2}.

\subsection{Extension 3: levels of stratification required for Shape Theory}

\n Appendix F's conceptual classification indicates, firstly, that the current paper's case of stratification in Shape Theory is particularly mild. 
We should look not to $N$ particles in 2-$d$ but to the harder case of $N$ particles in $\geq$ 3-$d$ as a source of further stratified examples. 
 
\m 

\n Secondly, in contrast, that \c{GT09, KKH16}'s affine and projective cases of stratification in Shape Theory are particularly strong.

\m 

\n These are strong and mild with respect to some established and highly controllable mathematical techniques developed: the middle case in our conceptual classification.   
This middle case is ready for use in global investigations, including global Background Independence and Problem of Time investigations \cite{ABook}.
On the other hand, the affine and projective's cases indicate that currently available methods fall short of being able to consider 
some of the more natural basic models for space (see \cite{Project-1, Project-2} for more detailed discussion). 

\m 

\n{\bf Acknowledgments} I thank Chris Isham and Don Page for previous discussions.  
Reza Tavakol, Malcolm MacCallum, Enrique Alvarez and Jeremy Butterfield for support with my career.  
I thank a particularly kind friend: I appreciate your style.  

\vspace{10in}

\begin{appendices}

\section{Outline of Order Theory and Lattice Theory}

\n{\bf Structure 1} A {\it binary relation} $R$ on a set $\FrX$ is a property that each pair of elements of $\FrX$ may or may not possess.  
We use $a \, R \, b$ to denote `$a$ and $b$ $\in \FrX$ are related by $R$'. 

\mbox{ }

\n{\bf Structure 2} Some basic properties that a binary relation $R$ on $\FrX$ might possess are as follows (i$\forall \, a, b, c \in \FrX$ wherever applicable).  

\m 

\n i) {\it Reflexivity}:  $a \, R \, a$. 

\m 

\n ii) {\it Antisymmetry}: $a \, R \, b$ and $b \, R \, a \Rightarrow a = b$.      

\m

\n iii) {\it Transitivity}: $a \, R \, b$ and $b \, R \, c \Rightarrow a \, R \, c$.  

\m 

\n iv) {\it Totality}: that one or both of $a \, R \, b$ or $b \, R \, a$ holds, i.e.\ all pairs are related.  

\mbox{ }  

\n{\bf Definition 1} A binary relation $R$ is a {\it partial ordering}, which we denote by $\preceq$, if $R$ is reflexive, antisymmetric and transitive. 
$R$ is moreover a {\it total ordering} alias {\it chain} if it is both a partial order and total.  

\m

\n Example 0) $\leq$ acting on the real numbers is a total ordering, whereas $\subseteq$ acting on sets as `is a subset of' is a partial ordering. 

\m

\n{\bf Definition 1} A {\it poset} is a set equipped with a partial order. is a {\it poset} $\langle \FrX, \preceq \rangle$.  

\m

\n{\bf Remark 1} (Small finite) posets are conveniently represented by Hasse diagrams \cite{Pure-Lattice}; see Figs 4, 8, 16, 20 for some examples of these.

\m

\n{\bf Definition 2} A {\it lattice} $\lattice$ \cite{Pure-Lattice, StanleyBook} 
is a poset for which each pair of elements possesses a {\it join} $\lor$ (least upper bound) and a {\it meet} greatest lower bound $\land$.

\m 

\n The study of orders constitutes {\it Order Theory} \cite{Pure-Lattice, StanleyBook},  
with {\it Lattice Theory} the itself-rich study of the specialization of this to lattices.

\m 

\n{\bf Example 1} A familar example of these operations is $\lor = \mO\mR$ and $\land = \mA\mN\mD$ in basic logic/computer science/electronics' formation of truth tables.

\m 

\n{\bf Proposition 1} $\lor$ and $\land$ form an algebra. 
More specifically, 

\m

\n i) each of these operations is commutative, associative, and idempotent.  

\m 

\n ii) They obey the {\it absorption conditions} 
\be 
a \lor \{ a \land b\} = a \m  \m{ and }   \m 
a \land \{a \lor b\} = a                  \m .
\ee 
[For Example 1), OR and AND form the {\it Boolean algebra}, which has further structure due to the presence also of a negation operation, by which NOR and NAND also enter.]

\m 

\n{\bf Proposition 2} $\lor$ and $\land$ are moreover dual operations. 

\m 

\n{\bf Definition 3} The {\it dual} of a given lattice is another lattice in which $\lor$ and $\land$'s statuses are reversed.  
The corresponding Hasse diagrams are upside-down relative to each other. 
If the arrows are furthermore reversed, the {\it antitone dual} is realized.  

\m 

\n{\bf Definition 4} An element 1 of $\lattice$ is a {\it unit element}         if 
\be 
\forall \, l \,  \in \,  \lattice   \mma 
l \preceq 1                         \m .
\ee 
An element 0 of $\lattice$ is a {\it null} or {\it zero element} if 
\be 
\forall \, l \,  \in \,  \lattice   \mma 
0 \preceq l                         \m . 
\ee
A lattice possessing both of these is termed a {\it bounded lattice}.

\m 

\n{\bf Definition 5} A {\it lattice morphism} is an order-, join- and meet-preserving bijection between lattices.  

\m

\n{\bf Example 2} The set of subsets of a fixed finite set $\FrX$ forms a bounded lattice $\lattice(\FrX)$ under the ordering `is a subset of'.   
The top and bottom elements here are $\FrX$ and $\emptyset$, and the join is the smallest subspace containing a pair of spaces. 

\m 

\n{\bf Example 3} The subgroups of a group form a bounded lattice $\lattice(\lFrg)$ under the ordering `is a subgroup of'.  
The top and bottom elements here are the whole group $\lFrg$ and the trivial group $id$, and the join is the subgroup generated by their union.

\m 

\n{\bf Example 4} Given an object space $\FrO\mb$ of objects $Ob$ that a group $\lFrg$ acts upon, some of the subgroups of $\lFrg$ may act identically. 
In this case, a smaller bounded lattice can be formed, of {\it distinct subgroup actions} on 
\be 
\FrO\mb, \lattice(\s{\rightarrow}{\lFrg} \FrO\mb) \m .
\ee    
Its top and bottom elements are the whole group's action on $\FrO\mb$ and the trivial action on $\FrO\mb$ (i.e.\ $id$ acting on each $Ob$ to simply return that $Ob$ again).  

\m 

\n{\bf Example 5} A subexample of the previous is for $\lFrg$ acting on $\FrO\mb$ to produce a quotient, 
by which the quotients under distinctly-acting subgroups themselves form a bounded lattice, 
\be 
\lattice(\frac{\FrO}{\Gamma}, {\Gamma} \in \lattice(\s{\rightarrow}{\lFrg} \FrO\mb)) \m . 
\ee  
The top element here is $Ob$ itself: the least quotiented space, whereas the bottom element is 
\be
\frac{\FrO\mb}{\lFrg}  \m : 
\ee 
the most quotiented space. 
[For configuration spaces or phase spaces, one often says `reduced' rather than `quotiented'.]
This lattice, as presented, is the antitone dual of $\lattice(\s{\rightarrow}{\lFrg} \FrO\mb)$.  

\m

\n{\bf Example 6} The lattice of monopoles is then lattice-isomorphic to the lattice of relational spaces: 
a subcase of quotient spaces of configuration spaces as defined in Sec 2.
The current paper's main output is two such bounded lattices: the bounded lattices of 2-$d$ and 3-$d$ monopoles for the 3-body problem under variable modelling conditions.  

\m 

\n{\bf Example 7} Order Theory underlies many advanced topics in Group Theory \cite{Group-Order}.  
Some of these have counterparts for more general structures -- algebras -- which enter Theoretical Physics for instance as 
classical-and-quantum constraint algebras and observables algebras and quantum-level operator algebras.
For instance, notions of observables form a lattice \cite{ABook, ABeables3} in a manner that depends on the lattice of subalgebras of the constraint algebra.  
In turn, gravitational theories necessitate further generalization to a constraint algebroid version of this.  

\m 

\n{\bf Remark 2} Order Theory is moreover the natural home for coarse-graining operations. 
While all physicists know some examples of coarse-graining, especially in phase space and classical and quantum statistical mechanics contexts, relatively few are aware of 

\m 

\n a) configuration space coarse-graining. 

\m 

\n b) That coarse-graining applies to a very wide range of mathematical structures, so in particular can be widely used in peeling away layers of structure one by one. 
E.g.\ purely topological features within a Differential Geometry modelling situation, 
or which of those purely topological features are topoligical space features rather than topological manifold features, 
or as regards which topological space features are in turn mererly set-theoretic.  
b) provides both of the following. 

\m

\n i) Useful layer-by-layer structural analysis: the current paper's use, separating out each of topological-shape and metric-shape's topological-shape-space 
features from metric-shape's metric shape-space features.  

\m 

\n ii) Alternative theories in which more or less layers of mathematical structure are dynamical and/or modelled in a Background Independent manner and/or quantized.   
See \c{I89-1, I89-2, I89-3, IKR-1, IKR-2, ASoS} and Epilogues II.C and III.C of \c{ABook} for examples of this at the topological, metric-space and underlying point-set levels.  

\vspace{10in}

\section{The paper's specific supporting Group Theory}

{\bf Remark 1} The permutation group on 3 distinct objects $S_3$ has elements 
\be 
\{1, \epsilon, \rho, \rho \, \epsilon, \, \rho^2, \, \rho^2\epsilon \} \mbox{ } , 
\ee 
where $\epsilon$ is the signature permutation and $\rho$ is an even permutation (for order 3, a 3-cycle). 
$S_3$ is furthermore isomorphic to the dihaedral group $D_3$, for which $\rho$ is a $\frac{2 \, \pi}{3}$-rotation and $\epsilon$ is a reflection. 
$S_3 \cong D_3$ additionally admits the standard `generators and relations' presentation 
\be
\langle \rho, \, \epsilon \, | \, \epsilon^2 = 1 = \rho^3, \, \epsilon \, \rho = \rho^{-1}\epsilon \rangle  \m . 
\ee
{\bf Remark 2} $S_3$ has 3 order-2 proper subgroups,  
\be 
C_2^1  :=  \{1 , \,         \epsilon\}  \mma 
C_2^2  :=  \{1 , \, \rho \, \epsilon\}  \mma 
C_2^3  :=  \{1 , \, \rho^2  \epsilon\}  \mma 
\l{Z1-3}
\ee 
and a single order-3 one, 
\be 
C_3 = \{1, \, \rho, \, \rho^2 \}   \m .
\l{Z3}   
\ee
{\bf Remark 3} Together with the full group $S_3$ and the trivial group $id = \{ 1 \}$, these form the bounded lattice of 6 subgroups
\be 
\lattice  \left( \mbox{ subgroups of } S_3 \m \right)  \m \mbox{ of Fig {S3-Z2-Lattice}.a) } \m , 
\ee 
whose arbitrary element we denote by $\overline{\Delta}$.  
As for any lattice of subgroups, the full group is the top element, whereas the trivial group is the bottom element.  

\m

\n{\bf Remark 4} Acting on the 3-$d$ shape hemisphere, 
the three $C_2$ subgroups have isomorphic group actions, corresponding to taking any pair of labels to be identical but different from the third.    
These form the bounded lattice 
\be
\lattice \left( \m   \mbox{ distinct } \s{\longrightarrow}{S_3  \mbox{ subgroup } \FrS(3, 2)} \m \right) \m \mbox{ square of Fig \ref{S3-Z2-Lattice}.b)} \m , 
\ee
whose arbitrary element we denote by $\Delta$.   

\m 

\n{\bf Remark 5} Acting on the claw graph of topological triangles, additionally the $C_3$ ceases to provide a nontrivial group action.
We are thus left with 
\be
\lattice( \m distinct \s{\longrightarrow}{S_3 \mbox{ subgroup }} \Top\mbox{-}\FrS(3, 2) \m ) = \mbox{ 3-chain } P_3 \mbox{ constituting Fig } \ref{S3-Z2-Lattice}.c) \m , 
\ee 
whose arbitrary element we denote by $\Delta^{\tau 2}$

\m
 
\n{\bf Remark 6} $S_3 \times C_2$ has, a priori, the direct product group presentation 
\be 
\{1, \epsilon, \rho, \rho \, \epsilon, \, \rho^2, \, \rho^2\epsilon \} \times \{1, \, \mu \}  \m ,  
\ee
for $\epsilon$ and $\rho$ as before and mirror symmetry $\mu$.  
The corresponding presentation of this, as directly implemented in the Shape Theory application in hand, is 
\be 
\langle \, \epsilon, \mu, \rho \, | \, \epsilon^2       = 1 = \mu^2        \mma 
                                     \rho^3         = 1                \mma 
				    				 \mu \, \rho    = \rho \, \mu      \mma 
								     \mu \, \epsilon  = \epsilon \, \mu    \mma  
                                     \epsilon \, \rho = \rho^{-1} \epsilon  \,   \rangle  \m .
\ee  
%
{            \begin{figure}[!ht]
\centering
\includegraphics[width=1.0\textwidth]{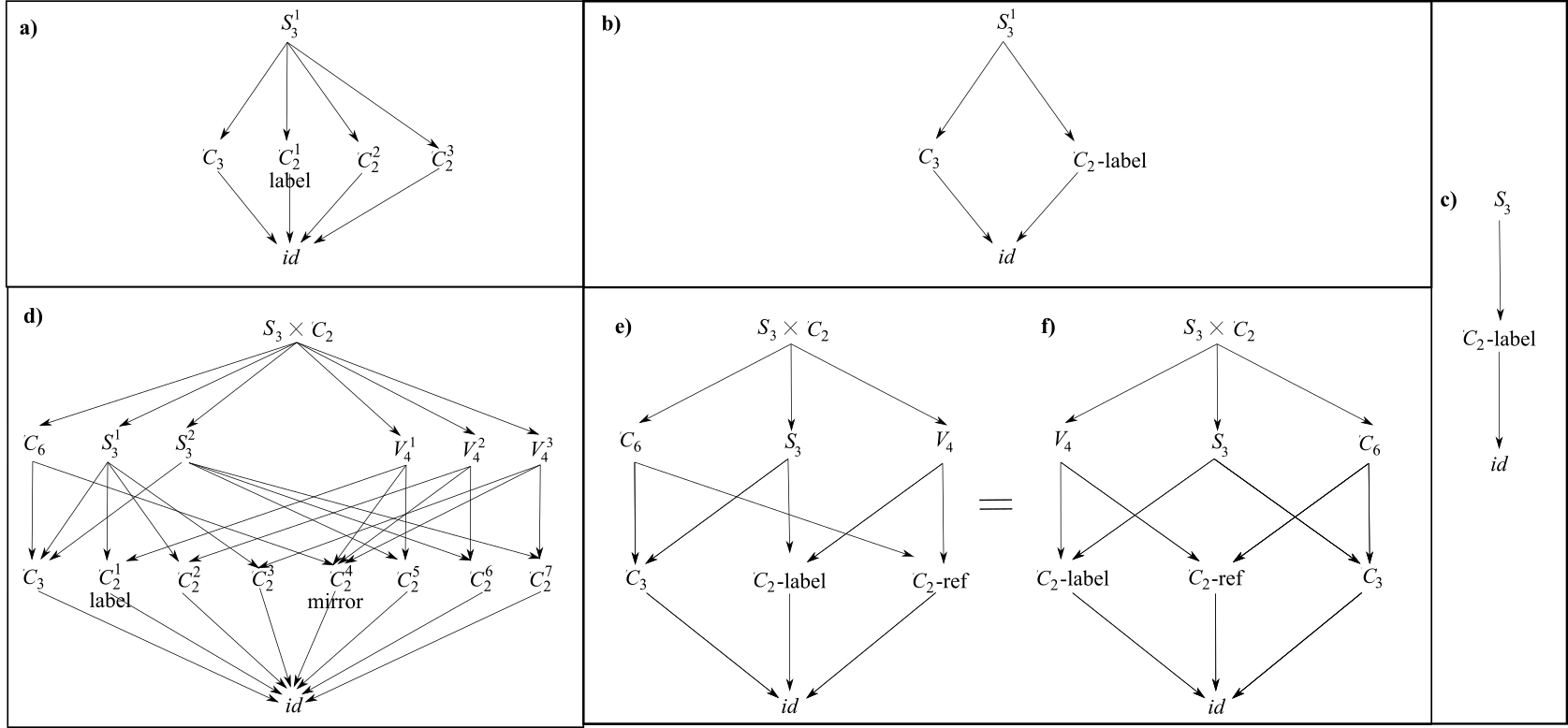}
\caption[Text der im Bilderverzeichnis auftaucht]{        \footnotesize{\n $S_3$'s lattices of 
a) subgroups and 
b) distinctly realized subgroup actions on the shape hemisphere of triangles in 3-$d$.
c) is the smaller lattice -- a chain -- of subgroup actions on the shape claw of topological triangles introduced in the next subsection.  

\m 

\n $S_3 \times C_2$'s lattices of 
d) subgroups and 
e) distinctly realized subgroup actions on the shape sphere of triangles in 2-$d$.
f) is the rearrangement of b) subsequently used in this paper. } }
\l{S3-Z2-Lattice} \end{figure}          }

\n{\bf Remark 7} While $S_3$ is already abstractly standard, it is not immediately clear which abstractly-standard order-12 group $S_3 \times C_2$ is isomorphic to.  
Order 12 moreover offers very few possibilities, and it is straightforward to check which of these is realized. 

\m 

\n{\bf Lemma 1} 
\be 
S_3 \times C_2  \cong D_6  \m .
\ee 
\Proof 
\be 
S_3 \times C_2  \cong  D_3                                  \times   C_2 
                         \cong  (C_3 \rtimes C_2)  \times   C_2 
						   =    (C_3 \times C_2)   \rtimes  C_2
						   =     C_6                        \rtimes  C_2 
						   =     D_6                                                          \m .  
\ee 
Here the first            step  uses $S_3 \cong D_3$: the dihaedral group of order 6. 
The      second and fifth steps use  the semidirect product structure of dihaedral groups. 
The      third            step can be computationally verified. 
The      fourth           step follows from the Classification Theorem for Abelian groups.          $\Box$ 

\m 

\n{\bf Remark 8} The standard $D_6$ generators and relations presentation is 
\be 
\langle \, x, \, y \, | \,  y^2 = 1 = x^6, \, y \, x = x^{-1}y  \, \rangle  \m . 
\ee 
Thus we identify the elements of the $S_3 \times C_2$ as directly implemented in the Shape Theory application in hand to be as follows. 

\m 

\n{\bf Lemma 2}
\be 
\rho    = x^2   \mma 
\mu     = x^3   \mma 
\epsilon  = y    \m , 
\ee 
\be 
\rho \, \mu  =  x^5        \mma
\rho^2  \mu  =  x^7  =  x  \mma 
\rho^2       =  x^4         \m , 
\ee 
\be 
\rho^2\epsilon \, \mu    =  x \, y  \mma 
\rho \, \epsilon         =  x^2  y  \mma 
\epsilon \, \mu          =  x^3  y  \mma
\rho^2\epsilon           =  x^4  y  \mma  
\rho \, \epsilon \, \mu  =  x^5  y   \m .
\ee 
\Proof Read off the first three by comparison of the $S_3 \times C_2$ and $D_6$ presentations.  
The other eight then follow. $\Box$

\m

\n{\bf Corollary 1} By Lemma 2, we can transcribe the standard knowledge of subgroups in the $D_6$ presentation to the $S_3 \times C_2$ group action as realized in Shape Theory. 
This gives the following suite of proper subgroups.   

\m 

\n There are 7 subgroups of order 2, $S_3$'s in (\r{Z1-3}), and
\be 
C_2^4 := \{1 , \, \mu\}                    \mma
C_2^5 := \{1 , \, \mu \, \epsilon\}          \mma
C_2^6 := \{1 , \, \rho \, \epsilon \, \mu\}  \mma
C_2^7 := \{1 , \, \rho^2\epsilon \, \mu\}    \m . 
\ee
\n There is just the one subgroup of order 3, coinciding with $S_3$'s in (\r{Z3}). 

\m
 
\n There are 3 subgroups of order 4 which are furthermore all Klein 4-groups, 
\be 
V_4^1 := \{1, \, \mu, \,           \epsilon, \,         \epsilon \, \mu \}  \mma 
V_4^2 := \{1, \, \mu, \, \rho \,   \epsilon, \, \rho \, \epsilon \, \mu \}  \mma 
V_4^3 := \{1, \, \mu, \, \rho^2 \, \epsilon  \, \rho^2  \epsilon \, \mu \}  \m .  
\ee 
\n Finally, there are 3 subgroups of order 6: one Abelian: 
\be
C_6 \mma \mbox{ realized as } \m \{1, \, \epsilon, \, \rho, \, \epsilon \, \rho, \, \rho^2, \, \epsilon \, \rho^2\} \m , 
\ee 
and two non-Abelian:  
\be 
S_3^1 := \{1, \, \rho, \, \rho^2, \, \epsilon, \, \epsilon \, \rho, \, \epsilon \, \rho^2  \}       \mma 
S_3^2 := \{1, \, \rho, \, \rho^2, \, \epsilon \, \mu, \rho \, \epsilon \, \mu, \, \rho^2 \epsilon \, \mu \}  \m .  
\ee 
\n{\bf Corollary 2} Together with the full group $S_3 \times C_2$ and the trivial group, these form the bounded lattice of 16 subgroups
\be 
\lattice  \left( \mbox{ subgroups of } \m S_3 \times C_2 \m \right) \mbox{ of Fig \r{S3-Z2-Lattice}.d) }  \m , 
\ee 
whose arbitrary element we denote by $\overline{\Gamma}$.  

\m

\n{\bf Remark 9} The various different possible models of partial indistinguishability and mirror image identification 
lie within what the above lattice of subgroups can support.  
This is not moreover a 1 : 1 correspondence, as various subgroups' {\sl group actions} on the triangleland shape sphere are moreover isomorphic, as follows.  

\m 

\n a) The three $V_4$'s act in isomorphic manner, as it makes no difference which two labels are allocated identical status. 

\m

\n b) By direct computation of which tiles are identified, $S_3^2$ is found to act isomorphically to $C_6$, whereas $S_3^1$ acts differently.

\m 

\n c)   The four $C_2$ subgroups that include $\mu$ -- labelled 4 to 7 -- act isomorphically, 
whereas the three which do not -- labelled 1 to 3 -- act in a distinct isomorphic manner. 

\m 

\n This leaves us with a total of $16 - 2 - 1 - 3 - 2 = 8$ distinct group actions, forming the lattice 
\be
\lattice  \left(  \mbox{ distinct } \s{\longrightarrow}{(S_3 \times C_2)} \FrS(3, 2)  \right) \m \m{cube of Fig \r{S3-Z2-Lattice}.f) } .
\ee
depicted in Fig \r{S3-Z2-Lattice}.e)-f) whose arbitrary element we denote by $\Gamma$.   
  
\m

\n{\bf Remark 10} Thus it is this eightfold which each produce a distinct quotient space, and consequently a distinct type of monopole.  
Due to this, the subsequent Figs \r{S(3, 2)-Top-8}, \r{S(3, 2)-Met-8} and \r{Monopoles-8} are underlied by anti-isomorphic lattices of quotient spaces and of monopoles.  
Anti-isomorphic means the same underlying lattice graph, but with the directions of the arrows reversed. 
We draw these upside-down relative to the original lattice of subgroup actions, for the reasons given in Fig \r{S(3, 2)-Top-8}.  

\m 

\n{\bf Remark 11} Acting on the claw graph of topological triangles, the new $C_2$ of mirror image identification has no nontrivial action. 
Thus the $S_3$ working for this is recovered: 
the lattice of distinct subgroup actions of $S_3 \times C_2$ on the shape claw is the just the same 3-chain as depicted in Fig \r{S3-Z2-Lattice}.c).

\section{Manifold-and-orbifold (chunk) notions and notation for the paper}
%
{            \begin{figure}[!ht]
\centering
\includegraphics[width=0.75\textwidth]{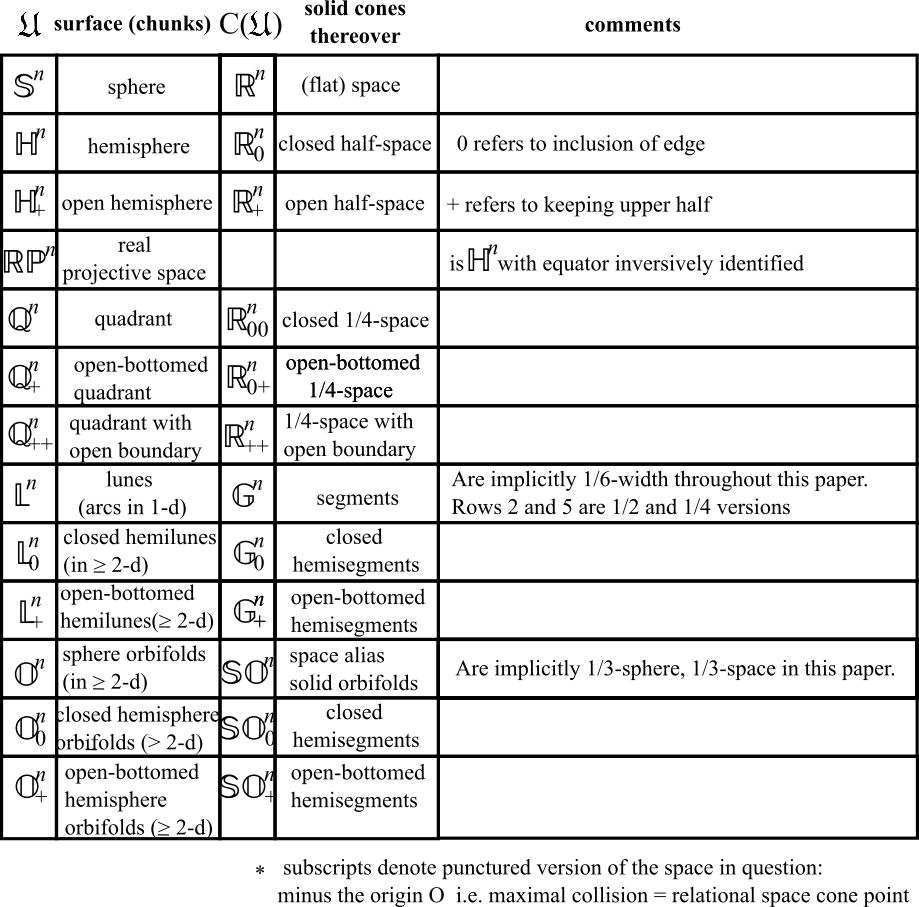}
\caption[Text der im Bilderverzeichnis auftaucht]{        \footnotesize{This paper's notation for specific manifolds and orbifolds, and for chunks thereof.
We subsequently effectively drop 1/6-labels on all lunes and 1/3-labels on all orbifolds.
Many of the 1- to 2- or 3-$d$ cases of these spaces are sketched within Figs \r{S(3, 2)-Top-8}, \r{Orbifold-Up} and \r{R(3, 2)-Met}; 
these other Figures are the main first points of application of this Appendix in the current Paper.} }
\l{Chunk} \end{figure}          }

\m 

\n We provide this in Fig \ref{Chunk}.  

\vspace{10in}

\section{Gauss--Bonnet(--Poincar\'{e}--Chern) Theorems}

\n{\bf Gauss--Bonnet Theorem} \c{NS83, Frankel, Berger} Let $\FrU$ be for now a closed orientable 2-surface equipped with a Riemannian metric, 
with $K$ the corresponding (Gaussian) curvature. 
Then  
\be
\int\int_{\sFrU} K \, \d A  \es  2 \, \pi \, \chi(\FrU)  \m . 
\l{GB}
\ee 
\n{\bf Remark 1} The left-hand-side is geometrical,  albeit global: the integral over the whole of $\FrU$.   

\m

\n{\bf Remark 2} In contrast, the right-hand-side is purely topological, proportional to the {\it Euler characteristic}  
\be 
\chi(\FrU) = 2(1 - g)  \m , 
\l{chi-g}
\ee 
for $\FrU$ of genus $g$.
In particular, 
\be 
\chi(\mathbb{S}^2) = 2 \m , 
\ee 
and, by means other than (\r{chi-g}), for the disc $\mathbb{D}^2$ 
\be 
\chi(\mathbb{D}^2) = 1.
\ee 
\n{\bf Remark 3} The Gauss--Bonnet Theorem's equatability of a priori distinct objects from different branches of Mathematics 
becomes the main feature in increasingly powerful Index Theorems \c{Nakahara, Nash, Shanahan}.

\m

\n{\bf Variant 1} For a 2-surface with boundary, the Gauss--Bonnet Theorem remains applicable if the surface integral term
\be
I_{\sg}  \es   \oint_{\pa \sFrV} \kappa_{\sg} \, \d s 
\ee 
is added on the left-hand side.
Here, $\pa S$ the component curves of the boundary and $\kappa_{\sg}$ is the {\it geodesic curvature} for the unit speed parametrizations of these curves.  
This integral $I_{\sg}$ itself is known as the {\it total geodesic curvature}.

\m

\n{\bf Remark 4} The current paper only makes use of one feature of the geodesic curvature: that it is zero for geodesic curves. 
This follows from 
\be 
0  \es  \frac{\mD \, W}{\mD \, t}  \m ,
\ee 
where $W$ is a differentiable unit vector field, $t$ is a parameter, and $\kappa_g$ is defined to be the numerical value of this covariant derivative.  
This case suffices because all of the current paper's boundaries are geodesics.  

\m

\n{\bf Variant 2} For a geodesic triangle, the right-hand-side term of (\r{GB}) also takes the form 
\be
\sum_{I = 1}^3 \alpha_I - \pi  \m , 
\ee
where $\alpha_I$ are the interior angles of the triangle.  

\m

\n{\bf Variant 3} For a geodesic $N$-a-gon, the right-hand-side term of (\r{GB}) takes the form 
\be
2 \, \pi \, \chi(\FrU) - E(\FrU)
\ee
for {\it exterior angle sum }
\be
E(\FrU) = \sum_{I = 1}^N \epsilon_I
\ee
and $\epsilon_I$ the exterior angles of the $N$-a-gon.

\m 

\n{\bf Variant 4} The theorem continues to hold for orbifolds $\FrO$, now with Euler characteristic calculated as follows. 
Let $\FrM$ be the $m$-fold covering manifold for $\FrO$.
Then we require    
\be
\chi(\FrO)  \es  \frac{\chi(\FrM)}{m}  \m . 
\l{chi-orb}
\ee 
\n{\bf Variant 5} ({\bf Gauss--Bonnet--Poincar\'{e} Theorem} \c{Frankel}) 
\be
\frac{  1  }{  2 \, \pi  } \int \int_{\sFrV} K \, \d  \, A  \es  \chi(\FrU) 
                                                        \es  \sum_{A} j_{v}(p_{A})  \m . 
\ee
Here $v$ is a vector field with singularities at $A = 1$ to $k$, and 
\be
j_v(p_A)  \:=  \frac{1}{2 \, \pi}\oint_{\pa \mathbb{D}^2} \d \psi \m 
\ee
is the {\it degree}: a simple notion of index, for $\psi$ the phase of $v$ and $\pa \mathbb{D}^2$ the boundary of a small disc excised around each singularity.  

\m

\n{\bf Variant 6} One can furthermore apply the Gauss--Bonnet Theorem to bundles \c{Isham, Nash, Husemoller}. 
Here the Chern curvature form \c{Frankel} takes on the role of the Gaussian curvature, giving the Gauss--Bonnet--Chern Theorem. 
See the next Appendix for the form taken by this in the case of Hopf-type bundles over (pieces of) the shape sphere.

\section{Hopf-type vector bundling over (pieces of) the shape sphere}
%
{            \begin{figure}[!ht]
\centering
\includegraphics[width=0.8\textwidth]{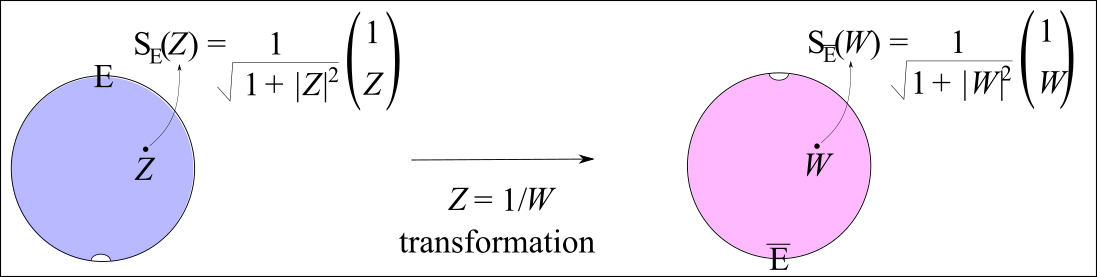}
\caption[Text der im Bilderverzeichnis auftaucht]{        \footnotesize{a) Hopf bundle. 
b) Its shape-theoretic realization. 
c)Two local sections for the Hopf bundle over $\mathbb{S}^2$, along with the inter-relating transformation.} }
\l{2-Local-Sections} \end{figure}          }

\n{\bf Structure 1} In the case free from discrete quotienting, the {\it Hopf vector bundle} is a principal fibre bundle, with 
\be 
\mbox{base space } \m \mathbb{S}^2  \m ,
\ee 
\be 
\mbox{fibre = structure group } \m \mathbb{S}^1 = U(1) = SO(2) = Rot(2)  \m ,
\ee 
and 
\be 
\mbox{total space } \m \mathbb{S}^3  \m .
\ee  

\m

\n In its shape-theoretic realization, this base space is the shape sphere $\FrS(3, 2)$, 
these fibres are are the space of representative triangles, covering the variety of absolute orientations \c{III}, 
and this total space is Kendall's preshape space $\FrP(3, 2)$.  

\m 

\n{\bf Structure 2} Start with the 2-chart set-up in Fig \r{2-Local-Sections}.  
Use 
\be
Z  :=  {\cal R} \, \mbox{exp} (i \, \Phi)         \m , 
\ee 
and 
\be
W  \es  Z^{-1} 
   \es  {\cal R}^{-1} \, \mbox{exp} (-i \, \Phi)  \m , 
\ee 
we deduce that
\be 
S_{\overline{\sE}}(W)  \es   \frac{  W  }{         \sqrt{1 + |W|^2}   }             \left(  \s{\mbox{1}}{W^{-1}}  \right)
                       \es   \frac{  Z^{-1}  }{    \sqrt{1 + \frac{1}{|Z|^2}  }  }  \left(  \s{\mbox{1}}{Z}       \right)
		               \es   \frac{|Z|}{Z} \, \frac{  1  }{         \sqrt{1 + |Z|^2}   }             \left(  \s{\mbox{1}}{Z}       \right)  
		               \es   t_{\sE \overline{\sE}}    \,   S_{\sE}(Z)                                                                       \m ,   
\ee 
for transition functions 
\be
t_{\sE\overline{\sE}} = \mbox{exp}(i \, \Phi)  \m : 
\ee 
a phase, which in this setting is the `Swiss-army-knife' relative angle (\r{Swiss}).   
Continuity of this phase gives a `topological quantization condition'; for the Dirac monopole in the spatial setting, this is the well-known\footnote{Here  
$e$ is the minimum quantum of electric charge, 
$\hbar$ is Planck's constant, and 
$g$ is the thereby also quantized minimum magnetic monopole charge. 
In the 3-body problem setting, $e$ and $g$ receive a different physical interpretation, though $g$ remains mathematically a {\sl monopole strength}.}  
   
\be 
\frac{2 \, e \, g}{\hbar}  \es  - 1  \m .  
\ee 
For the orbifold monopole of Case 3), the left-hand-side is $-\frac{1}{3}$, 
so the monopole strength is $\frac{1}{3}$ that of the configuration space realization of the Dirac monopole in Case 1). 

\m

\n{\bf Remark 1} For the other ten cases in the current paper, moreover, a single chart suffices rather than needing to define two and a transition function between them.
Thus these do not elicit topological quantization conditions.  

\m

\n{\bf Remark 2} Case 3) moreover involves an {\sl orbibundle} version of the Hopf bundle; see \c{TF17} for a review of orbibundles. 
%

\m 

\n{\bf Structure 2} The corresponding {\it Simon connection} \c{Simon, Frankel} is, now specifically in the shape-theoretic context, the 1-form  
\be 
\omega_{\sE}(Z)  \es  \frac{i \, {\cal R}^2 \, \d \Phi}{1 + {\cal R}^2}  \m . 
\ee 
\n{\bf Structure 3} The corresponding curvature form is [$U(1)$ is of course Abelian] 
$$ 
\theta_{\sE}(Z)  =                           \d \omega_{\sE}(Z) 
                \es   \d \, \frac{  i \, {\cal R}^2 \, \d \Phi  }{  1 + {\cal R}^2  } 
		        \es    i \left\{   \frac{  2 \, {\cal R} \, \d {\cal R}  }{  1 + {\cal R}^2  }                     \m - \m 
		                           \frac{  {\cal R}^2  }{  \{1 + {\cal R}^2\}^2} \, 2 \, {\cal R} \d {\cal R}      \m + \m
		                           \frac{  {\cal R}^2 \, \d^2 \Phi  }{  1 + {\cal R}^2  }                                            		   \right\}
$$  
\be 
                \es    i \,      \frac{  2 \, {\cal R} \, \d {\cal R} \{ - {\cal R}^2 + 1 + {\cal R}^2  \}  }{    \{1 + {\cal R}^2\}^2    }  
		        \es    2 \, i \, \frac{       {\cal R} \, \d {\cal R} \wedge \d \Phi      }{      \{1 + {\cal R}^2\}^2      }                             \m . 
\l{thetaU}
\ee 
Here we used $\d^2 = 0$ in the fourth equality.  

\m

\n The {\it Chern curvature form} of the shape-theoretic Hopf bundle is then 
\be 
C_{\sE}(Z)  \:=  \frac{  i \, \theta_{\sE}(Z)  }{  2 \, \pi  }  \m .
\ee 
\n{\bf Remark 2} For the sphere, we can check that this obeys
\be 
\int\int_{\mathbb{S}^2} \biC  \es  - \frac{1}{4 \, \pi} \int\int_{\mathbb{S}^2} K \, \d A  \m . 
\ee 
This gives the {\it Gauss--Bonnet--Chern} formulation of the Gauss--Bonnet Theorem   
\be 
G(\FrU)  \es  T(\FrU) \m ,
\ee 
for {\it total geometrical contribution}   
\be 
G(\FrU)  \es  \int\int_{\sFrU} \biC
\l{G(S)}
\ee
and {\it total topological contribution}   
\be 
T(\FrU)  \:=  \frac{1}{2} \, \left\{ \, {E(\FrU)}{2 \, \pi} - \chi(\FrU) \, \right\}  \m .  
\l{T(S)}
\ee
\n{\bf Remark 3} To cover the entirety of this paper, then, $\FrU$ is to have the status of a {\it (possibly stratified) manifold-or-orbifold chunk}. 
This (or some technical restriction thereof) is Shape Theory's analogue of Algebraic Geometry's notion of variety ${\FrV}$ \cite{Hartshorne}.  
For the current Paper, as Appendix F explains, the notion of trivially contiguous stratified manifold-or-orbifold (chunk) suffices, 
and at least for the current paper's examples, Gauss--Bonnet--Chern immediately extends to these.  

\mbox{ }

\n{\bf Remark 3} As regards integrating over the whole (shape) sphere, the 
\be
{\cal Y} = {\cal R}^2
\l{X=R2}
\ee 
substitution will do: 
$$
G(\mathbb{S}^2)               \es   \int\int_{\mathbb{S}^2} \biC  \es   \frac{i}{2 \, \pi} \int\int_{\mathbb{S}^2} \theta_U 
                              \es   \frac{i}{2 \, \pi} \int_{\Phi = 0}^{2 \, \pi} \d \Phi   
						                              \int_{{\cal R} = 0}^{\infty} \frac{  i \, 2 \, {\cal R} \, \d {\cal R}  }{  \{1 + {\cal R}^2\}^2  } 
						      \es  -\frac{1}{2 \, \pi}	\, 2 \, \pi \int_{{\cal Y} = 0}^{\infty} \frac{\d {\cal Y}}{\{1 + {\cal Y}\}^2} 
$$
\be 						  
		   				      \es   \left[  \frac{1}{1 + {\cal Y}}  \right]_{{\cal Y} = 0}^{\infty} 
						      \es   \frac{1}{\infty}  \m - \m  \frac{1}{1 + 0} 
						      \es   0 - 1 
						       =    - 1                                                                                                                     \m .  
\ee
Here the first equality makes use that the deleted point precluded from the chart $\mE$ contributes zero measure to the integral. 
The second uses (\r{thetaU}), and the third (\r{X=R2}).

\m

\n{\bf Remark 4} For geodesic pieces of the shape sphere, as the rest of the paper considers, the substitution (\r{RTheta}) (especially for the usual $\Theta$ with pole $\mE$). 
This is recognized as the shape-theoretic version of the stereographic radius to azimuthal spherical coordinate transformation.
Here,  
\be
\frac{4 \, {\cal R} \d {\cal R} \, \d\Phi}{\{1 + {\cal R}^2\}^2}  \es  \, 2 \, \sin \, \frac{\Theta}{2}\, \mbox{cos}\,\frac{\Theta}{2} \, \d\Phi     \, \d\Theta   
                                                                  \es          \sin \, \Theta                                          \, \d\Phi     \, \d\Theta                                            
																   =           \sin \, \w{\Theta}                                      \, \d\w{\Phi} \, \d\w{\Theta}      \m ,  
\ee 
where the last step is merely a $\pi/2$ rotation of coordinates. 
Thus 
\be 
\theta_{\sE}  \es  \frac{i}{2} \, \sin \,    \Theta  \, \d    \Phi  \wedge \d    \Theta
              \es  \frac{i}{2} \, \sin \, \w{\Theta} \, \d \w{\Phi} \wedge \d \w{\Theta}  \m .
\ee 
This gives 
\be 
G(\FrU)  \es  - \frac{1}{4\pi}\int\int_{\sFrU} d A 
         \es  - (\mbox{ proportion of shape sphere included in $\FrU$ }) \m .  
\ee

\vspace{10in}

\section{Conceptual classification of types of stratification arising in Shape Theory}

Manifolds are in general insufficient for the purpose of studying reduced configuration spaces $\FrQ/\lFrg$.
This quotienting more generally produces unions of manifolds whose dimensions in general differ.
Thus there is a breakdown at least of the `locally Euclidean' pillar of the theory of manifolds (the other two being Hausdorffness and second-countability).  

\m   

In the current paper's examples (and elsewhere in Shape Theory, so far), these pieces moreover `fit together' according to some fairly benevolent rules 
put forward by Whitney \cite{Whitney46, Whitney65} and Thom \cite{Thom55, Thom69}. 

\m

\n{\bf Structure 1} Let $\FrX$ be a topological space that is not presupposed to be a topological manifold.
Suppose a topological space $\FrX$ can be split according to $\FrX = \FrX_{\sp} \bigcup \FrX_{\sq}$. 
Here $\FrX_{\sp} := \{\mp \in \FrX, \mp \mbox{ simple}\}$, $\mbox{dim}_{\sp}(\FrX) = \mbox{dim}(\FrX)$ 
where `simple' means `regular' and `ordinary', and $\FrX_{\sq} := \FrX - \FrX_{\sp}$.  
Consider such splittings recursively, so e.g.\ $\FrX_{\sq}$ further splits into $\{\FrX_{\sq}\}_{\sp}$ and $\{\FrX_{\sq}\}_{\sq}$.
Then setting $\FrM_1 = \FrX_{\sp}$, $\FrM_2 = \{\FrX_{\sq}\}_{\sp}$, $\FrM_3 = \{\{\FrX_{\sq}\}_{\sq}\}_{\sp}$ etc.\ gives 
             $\FrX = \FrM_1 \bigcup \FrM_2 \bigcup  \, ... \,  , \mbox{dim}(\FrX) = \mbox{dim}(\FrM_1) > \mbox{dim}(\FrM_2) > \, ...$, 
			 where each $\FrM_{\sfI}$ $\fI = 1, 2, \, ...$ is itself a manifold.
 
\m 
 
\n{\bf Remark 1} This procedure partitions $\FrX$ by dimension. 
$\FrX$ is moreover only a topological manifold if this is a trivial (i.e.\ single-piece) partition.

\m 

\n{\bf Definition 1} On the other hand, a {\it strict} partition of a topological space is a (locally finite) partition into strict manifolds. 
[A manifold $\FrM$ within a $m$-dimensional open set $\FrW$ is $\FrW${\it -strict} if its $\FrW$-{\it closure} 
$\overline{\FrM}$ := $\FrW - \mbox{Clos}\,\FrM$ and the $\FrW$-{\it frontier} $\overline{\FrM} - \FrM$ are topological spaces in $\FrW$.]  

\m 

\n{\bf Definition 2} A set of manifolds in $\FrW$ has the {\it frontier property} if, for any two distinct such, say $\FrM$ and $\FrM^{\prime}$,  
\beq
\mbox{ if } \FrM^{\prime} \bigcap \overline{\FrM} \neq \emptyset \mbox{ } , \mbox{ } 
\mbox{ then } \FrM^{\prime} \subset \overline{\FrM}              \mbox{ } \mbox{ and } \mbox{ } \mbox{ } 
                  \mbox{dim}(\FrM^{\prime}) < \mbox{dim}({\FrM}) \mbox{ } .
\label{Frontier}
\eeq
A partition into manifolds is itself said to have the frontier property if the corresponding set of manifolds does. 

\m 

\n{\bf Definition 3} A {\it stratification} of $\FrX$ \cite{Whitney65} is a strict partition of $\FrX$ that possesses the frontier property. 
The corresponding set of manifolds are known as the {\it strata} of the partition.

\m

\n{\bf Remark 2} Stratified manifolds have additionally been equipped with differentiable structure (see e.g.\ \cite{Sniatycki, Pflaum}).

\m 

\n{\bf Remark 3} All of the above can be carried over to the case of orbifolds as well, yielding stratified orbifolds \cite{TF17}.  

\m

\n{\bf Type i) Trivially-contiguous}. 

\m 

\n These stratified spaces can be qualified as manifolds-or-orbifolds with boundaries, corners (etc.\ in higher dimensions), 
in which some of the boundaries, corners etc are geometrically distinct, but are still contiguous to the top stratum in the manner of manifold theory.

\m 

\n Example 1) The cone over a compact manifold can be viewed as a stratified orbifold, with the apex and the remainder as strata; contiguity is clear: 
the remainder `wraps around' the apex.  
In the case in which the angle around the apex is the usual one, a stratified {\sl manifold} description suffices.  

\m 

\n Example 2) The current paper's strata are clearly trivially contiguous. 
Some form cones, with their most troublesome stratum -- the maximal collision -- forming the cone point alias apex.  
At the topological level, moreover, the relational space's non-collision collinearities become indistinguishable from generic triangular configurations. 
But our 3-$d$ spaces' stratification remains a feature of the binary collisions -- themselves topologically distinct -- as well as of the maximal collision. 
\n See Fig \r{Strat-Ex} for a notation summary table of this paper's 8 examples of trivially-contiguous statified manifolds.
%
%
{            \begin{figure}[!ht]
\centering
\includegraphics[width=0.6\textwidth]{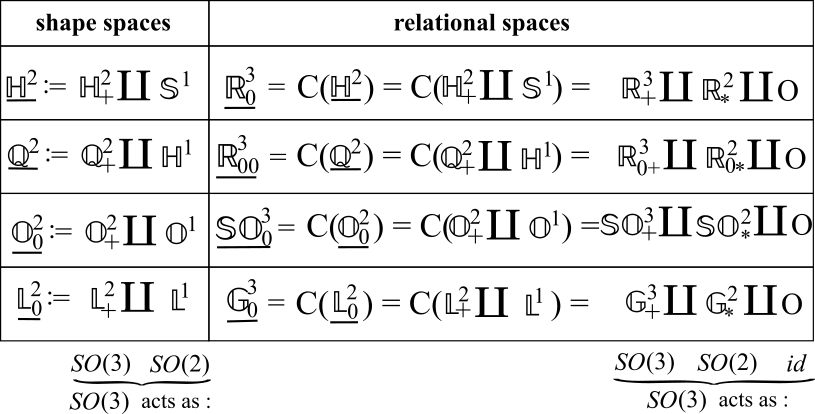}
\caption[Text der im Bilderverzeichnis auftaucht]{        \footnotesize{This paper's notation for specific stratified manifolds and orbifolds.  
This uses $\w{\Phi}$ as equatorial angle; 
the 3 particles in 1-$d$ point of view \cite{I} prefers to use a half-angle version of this, $\varphi = \w{\Phi}/2$; 
our choice in this regard is to maximally simplify the presentation of the strata.                                     } }
\l{Strat-Ex} \end{figure}          }

\m

\n{\bf Type ii) Topologically-nice and stratificationally-contiguous}.  

\m 

\n Here we require 

\m 

\n a)  a nontrivial realization of the stratification contiguity condition 

\m 

\n b)  That $\FrX$ be either LCHS (locally compact Hausdorff second-countable) \c{Lee}. 
In this case, the other two pillars of manifoldness are kept and moreover supplemented with a further analytic niceness condition, local compactness:  
that each point $\mp \in \FrX$ is contained in a compact neighbourhood.    

\m 

\n Or that $\FrX$ be LCHP (locally compact Hausdorff paracompact) \cite{Munkres}; 
there is moreover considerable degeneracy between paracompactness and second-countability in the current context.

\m 

\n Stratified manifolds and fibre bundles do not moreover fit well together due to stratified manifolds' local structure varying from point to point.  
Three distinct strategies to deal with this are outlined in Epilogue II.B of \cite{ABook} 
Among these, relational considerations point to the strategy of accepting the stratified manifold.  
In turn, this points to seeking a generalization of Fibre Bundle Theory, for which Sheaf Theory \cite{Sheaves, Sheaves1} is a strong candidate.  

\m 
 
\n Conceptual and computational schemes have been provided for stratified manifolds which are LCHS by Kreck \cite{Kreck} and LCHP by Pflaum \cite{Pflaum}.
These approaches in good part involve pairing stratified manifolds with sheaves; in Kreck's case, this is termed a {\it stratifold}.  

\m 

\n{\bf Type iii) Topologically-complex stratificationally-contiguous}. 
Affine and projective Shape Theory \c{GT09, KKH16} necessitate consideration of merely Kolmogorov-separated stratified spaces .  

\m

\n{\bf Example 3} the Affine Shape Theory of quadrilaterals in the plane, both the collinear and generic shapes form their own real projective space $\mathbb{RP}^2$ stratum, 
\be 
\mathbb{RP}^2 \, \coprod \, \mathbb{RP}^2 \m , 
\ee 
with every collinear configuration C lying arbirtarily close to every generic configuration G.
This impossibilitates Hausdorff separability, or even Fr\'{e}chet seprability \cite{Willard}; 
all that one is left with is the much weaker and {\sl qualitatively distinct} Kolmogorov separability \cite{Willard}. 

\m

\n{\bf Remark 4} Theorems of Analysis are more sparsely available here, and computational schemes for stratified manifolds of this more general nature remain to be developed.

\end{appendices}

\vspace{10in}


\end{document}